\documentclass{article}
\usepackage{graphicx}

\usepackage[utf8]{inputenc}
\usepackage{amsmath,amssymb}
\usepackage[margin=1in]{geometry}
\usepackage{color}
\usepackage{ulem}
\usepackage{marginnote}
\usepackage{url}
\usepackage{authblk}
\usepackage{enumerate}
\usepackage{amsthm}

\title{Cold Extremes during Dansgaard-Oeschger Oscillations}

\author[1]{Ignacio del Amo \footnote{ignacio.delamo@nbi.ku.dk}}
\author[1]{Peter Ditlevsen}

\affil[1]{Niels Bohr Institute, University of Copenhagen, Copenhagen, DK 1165, Denmark}

\date{\today}

\begin{document}

\maketitle

\begin{abstract}
    This paper studies the statistics of extreme cold air surface temperatures in a climate that experiences abrupt changes. We employ a CCSM4 simulation of Last Glacial Maximum conditions that exhibits rapid switching between stadial and interstadial states as data. Non-stationary linear Generalized Extreme Value (GEV) distributions are fitted to find the regions that show the most prominent changes and associate them with physical processes. The results are then compared with a non-linear model, which gives a more detailed picture of how the parameters change as a function of the AMOC strength. While different regions of the world show different extremal behaviours, many regions show an approximately linear relationship between the parameters of the GEV distributions and the strength of the AMOC within the stadial and interstadial states, with some non-linear oscillation or jump where the transition between them takes place. 
    Physical processes such as the expansion and retreat of the sea ice and the relative changes in the strength of the currents are shown to impact the magnitude and variability of the extremes, with significant changes observed in the three parameters of the GEV. They also create teleconnections that are compared whenever possible with various proxies for the temperature of the air and the water. 
    We show how mapping the parameters of the GEV distributions into the AMOC strength gives a way to compare results between different climate models and different climate states. Comparisons, however, need to pay heed to the dynamical characteristics of the state and the location to be meaningful.
\end{abstract}

\section{Introduction}



Changes in the frequency and intensity of extreme events are amongst the most catastrophic and harmful effects of climate change \cite{Aghakouchak2013ExtremesIA}. Heatwaves are increasing in intensity and frequency, causing widespread excess mortality and wildfires in summer \cite{galfi2021fingerprinting}. Cold spells can cause agricultural disasters and migrations, and they are also changing behavior \cite{galfi2021fingerprinting}. Even if a warmer world generally reduces cold spell frequency and intensity \cite{huang2016estimating,galfi2021fingerprinting}, it also alters physical processes that can result in extreme conditions, such as the destabilization of the polar vortices \cite{screen2014arctic}.
A main driver of heat transport is the Atlantic Meridional Overturning Circulation (AMOC). A slowdown of the AMOC could cause a drop in temperatures globally and an increase in cold extremes, especially in the Northern Hemisphere \cite{meccia2024extreme,van2025european}. Here we investigate the dependency between cold extremes and AMOC strength using output from a climate model that displays Dansgaard-Oeschger-like oscillations in Last Glacial Maximum (LGM) conditions.

Statistical extremes of a distribution behave in a fundamentally different way than average quantities. The asymptotic distribution for the annual maxima or minima of a stationary variable is given by a Generalized Extreme Value (GEV) distribution \cite{leadbetter2012extremes,lucarini2016extremes}. The GEV distribution is defined by 3 different parameters. These parameters are the location parameter $\mu$, the scale $\sigma$ and the tail parameter $\xi$. Each parameter describes the distribution in a different way. The location parameter describes where the bulk of the distribution is, the scale parameter describes how broad the distribution is, and the tail parameter describes how fast the tails of the distribution decay to zero. While the location parameter and the scale parameter are akin to the mean and the variance of the distribution, they are not the same. If the tail parameter is big enough ($\xi\geq 1/2$), the distribution becomes fat tailed and variance does not converge. Similarly, for $\xi\geq 1$ the average does not exist either. The tail parameter governs how skewed the distribution is and shows how much distributions of extremes can depart from Gaussian statistics, which are sometimes used to model tails of distributions of extreme events \cite{Aghakouchak2013ExtremesIA,krakauer2024normal}. For these reasons, the tail parameter is considered the most important and describes a kind of behaviour not present in Gaussian statistics.

A problem arises when the variable under study is not stationary. Although there are some results of convergence to a GEV for extremes of variables that become stationary sufficiently fast \cite{freitas2017extreme}, non-stationarity, in general, prevents convergence. Assuming that the underlying processes generating the extremes are sufficiently fast compared to the drift of the system, an alternative approach is possible: To make the parameters of the GEV distribution functions of time, under the assumption that the extremes still follow approximately a GEV distribution, albeit as one that shifts \cite{Aghakouchak2013ExtremesIA,coles2001introduction,robin2020nonstationary,xavier2020stationary}.  

As climate change makes the distributions of extremes shift, taking non-stationarity into account in modeling becomes a necessity \cite{slater2021nonstationary}. This allows us to integrate new and past data into the analysis. The temporal shift in extreme statistics in climate data is, in many cases, due to the comparatively slow drift induced by anthropogenic warming in the fast dynamics of atmospheric variables such as temperature and precipitation \cite{robin2020nonstationary,xavier2020stationary}. But there are also other variables, for example in hydrology, for which the non-stationarity may come from other environmental changes, such as land use or water management, or take the shape of a point change, a change in the statistics brought about by a one time event \cite{slater2021nonstationary}.

In this paper, we extend the method of analysis from \cite{coles2001introduction} based on a covariate to a climate system that exhibits abrupt transitions between two different climate states. This type of change is fundamentally different from the slow drift investigated in other studies. From a dynamical perspective, it is known that changes in the attracting set of the dynamics are related to the changes in the parameters of the extreme value laws that describe extremes of observables in a dynamical system \cite{faranda2011numerical,lucarini2012extreme,lucarini2012universal}. Thus, it is possible to observe changes in the tail parameter, which are, in many cases, not expected or not thought to be reliable if the change is small because it is difficult to estimate the tail parameter accurately \cite{coles2001introduction,robin2020nonstationary}.

The system chosen to study is a low resolution ($\sim3^\circ \times 3^\circ$) CCSM4 simulation containing Dansgaard-Oeschger events presented in ref.\cite{vettoretti2022atmospheric}. This simulation has LGM conditions, but some variables, such as the atmospheric CO$_2$ concentrations and the land ice sheet sizes, are fixed; see ref.\cite{vettoretti2022atmospheric} for more information. In particular, we focus on the simulation that has 200 ppm of atmospheric CO$_2$ concentration. This simulation is almost 8000 years long, with monthly resolution and exhibits spontaneous oscillations between a stadial and interstadial climate. 

We study cold extremes, which are directly influenced by the switching between states and are a source of concern for future AMOC tipping scenarios. Previous work that has explored similar topics is \cite{van2025european}, in which a full hysteresis experiment is conducted and extreme value distributions for AMOC ON/OFF scenarios are computed. Other similar work is  \cite{meccia2024extreme} in which the impact of a reduced AMOC in cold spells is investigated, but from a dynamical perspective and without the use of Extreme Value Theory. Another study that has looked at changes in the parameters of GEV distributions is \cite{huang2016estimating}, in which they study how the temperature extremes change over the USA in an increase of CO$_2$ scenario.

Our approach is fundamentally different and relies on viewing the parameters of the GEV distribution themselves as observables conveying some physical meaning, which can be related to the underlying physical processes generating the extremes. From a statistical point of view, knowing the possible drivers of non-stationarity for the extremes is useful. Since assessing whether data is non-stationary or not can be misleading, a physical mechanism providing a plausible cause for non-stationarity is important evidence \cite{slater2021nonstationary}.

The paper is organized as follows. In the second section, we employ a linear model that assumes that the temperatures vary approximately linearly with the strength of the AMOC. This model is unrealistic but allows us to perform statistical tests for non-stationarity and detect areas where the changes in the parameters are significant. In the third section, we study how the parameters change against the AMOC in these areas without imposing a model. This shows a more realistic dependence and reveals many different behaviours that are realized simultaneously in the data. In the fourth section, we try to relate some of the teleconnections observed to physical processes and proxy records. In the fifth section, we try to envision how this information about the climate could be used to extrapolate to other climate states or models and discuss its limitations.




\section{Linear model for trend detection}

\begin{figure}
    \centering
    \includegraphics[width=\linewidth]{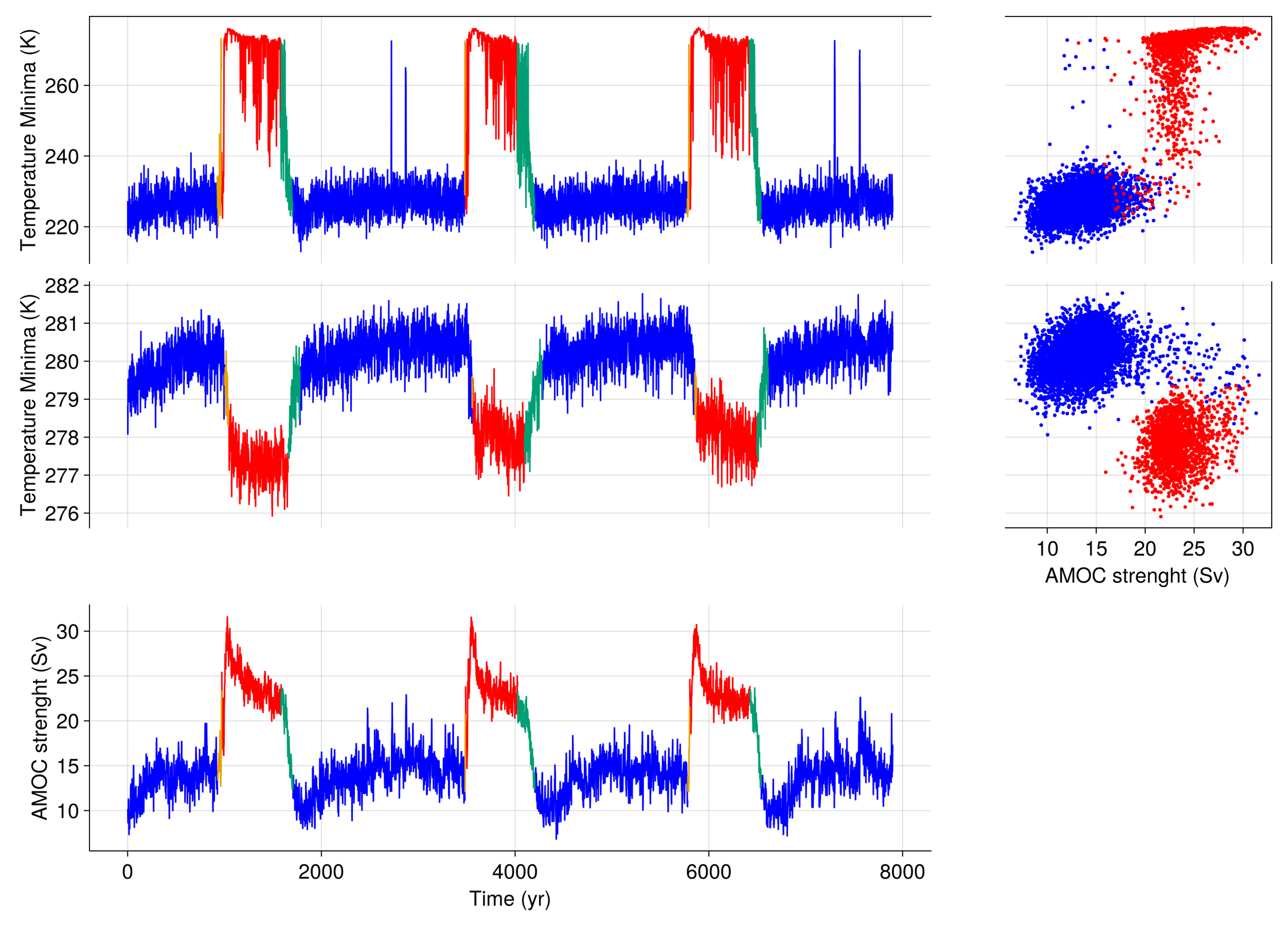}
    \caption{On the top, data series of the annual minimum temperature in the grid point 30$^\circ$W 57$^\circ$N. In the left panel it is plotted against time and in the right panel against the AMOC strength. The plots in the middle correspond to the location 34$^\circ$E 42$^\circ$S plotted in the same fashion. The bottom panel shows the AMOC strength plotted against time. The colour scheme corresponds to a division of the data series between stadial (blue), interstadial (red) and the transitions S$\rightarrow$I (yellow) and I$\rightarrow$S (green). In the right panels the data points corresponding to the transitions have been removed to give a more clear picture of the states.}
    \label{fig:division_figure}
\end{figure}

In this section, we use a linear model to see in which regions of the world significant patterns are found. Figure~\ref{fig:division_figure} shows how the data are visualized. The minimum annual temperature in two chosen locations is plotted against time in the top left panels and against the AMOC strength in the top right panels. Also, the AMOC strength is plotted against time in the bottom panel. 
The minima are colour-coded so that blue points correspond to the stadial state, red points correspond to the interstadial state, yellow corresponds to the transition from stadial to interstadial (S$\rightarrow$I), and green corresponds to the transition from interstadial to stadial (I$\rightarrow$S). Colouring of the data series is lag-corrected to maximize the absolute value of the correlation between the temperatures and the AMOC. The division of the data into stadial and interstadial has been done algorithmically (see Appendix~\ref{ap:division_algorithm}), and the points corresponding to the transitions are left out of the analysis in this paper. 

The oscillation of the AMOC strength is an important part of the D-O oscillations \cite{boers2018ocean}. Due to the feedbacks present between the temperatures, sea ice, and ocean currents \cite{vettoretti2022atmospheric}, the AMOC strength shows the same abrupt changes as the temperatures do while retaining a significant correlation with the temperatures within each state. This makes it an ideal candidate for the following model 

\begin{align}\label{eq:linear_model}
\begin{array}{c}
    \mu(t)= \mu_0+\mu_1\cdot\text{COV}(t)\\
    \sigma(t)=e^{\sigma_0+\sigma_1\cdot\text{COV}(t)}\\
    \xi(t)=\xi_0+\xi_1\cdot\text{COV}(t)
\end{array}
\quad \mathrm{COV}(t)= \mathrm{AMOC}(t)
\end{align}

This is a linear non-stationary GEV model. The model assumes that the parameters of the GEV distribution change as a linear function of a time dependent covariate, in this case, the AMOC strength $\mathrm{AMOC}(t)$. The parameter $\sigma$ is taken to be the exponential of a linear function to ensure that it remains greater than zero. 
Note that other variables are more significantly correlated with the minima (e.g., the average temperature in each location), and thus they linearize the data better. However, they can obscure the patterns that we are looking for because of the excess correlation that might reflect inter-annual variability in each location. (Using the global average temperature instead of the AMOC strength should produce very similar results since they are very strongly correlated $\approx0.856$).


An advantage of the linear model is that it allows for testing non-stationarity in each of the parameters \cite{coles2001introduction}. Non-stationarity may be present only in some parameters of the model, and the test thus serves as a tool to detect when a significant trend can be identified in any of the different parameters. 
This means that for each data series we have 8 possible models to fit, corresponding to all the possible choices of parameters that could be stationary. Thus, there is a stationary model, 3 models in which only one parameter changes, 3 models in which two parameters change, and the full non-stationary model where all parameters change over time. We have fitted all the models and performed pairwise tests whenever possible to find the best fitting model at a given significance level. The method is fully explained in Appendix~\ref{ap:linear_model}.

\subsection{Results}

\begin{figure}
    \centering
    \includegraphics[width=0.54\linewidth]{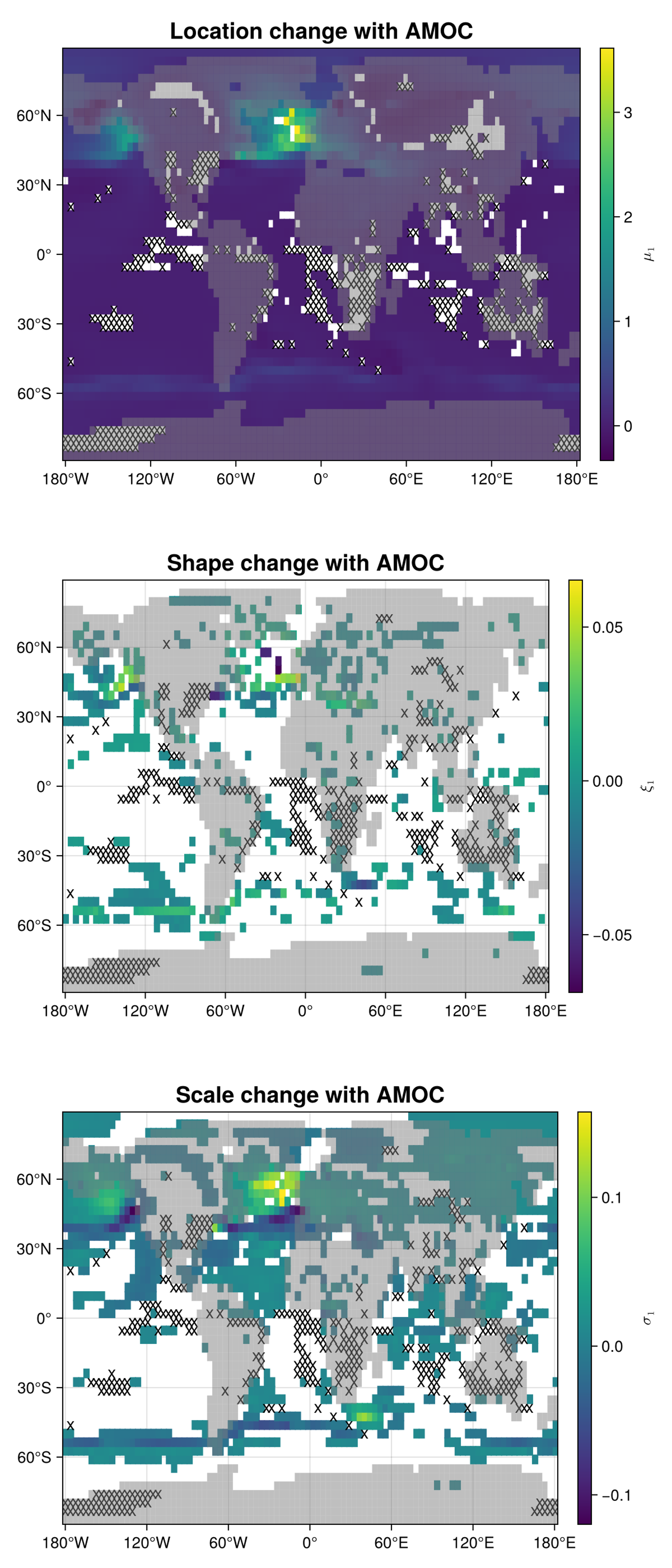}
    \caption{Sensitivity parameters of AMOC change. On the top $\mu_1$, on the middle $\sigma_1$ and in the bottom $\xi_1$. The crosses represent points where the chosen model is the stationary one $\mathcal{M}_\emptyset$ and the blank spaces points where the parameter value is zero but the model is non-stationary in some other parameter.}
    \label{fig:planetplot}
\end{figure}

Figure~\ref{fig:planetplot} shows the results of the statistical analysis at confidence level $99.999\%$. Note that the confidence level should not be interpreted as confidence that a linear relationship exists between the parameters of the GEV that fit data and the covariate, but rather as evidence that a linear relationship between them explains the temporal evolution of the data better than a stationary model in any given non-stationary parameter.

Trends are detected in almost every spatial dimension of the model. 
There are weak trends present almost everywhere except in some areas around the equator, in the Pacific Ocean, the centre/south of Africa, and the interior of Australia, Antarctica, and North America, where trends do not exist at the given confidence level. 

Additionally, there are areas that show a very strong trend. Notably, we see the strongest changes in the North Atlantic region. It is the area most strongly correlated with the strength of the AMOC, which 
is expected given its role as a major driver of heat transport in the area. Other regions that show significant correlations are the North Pacific area, specifically Alaska and the waters of the Gulf of Alaska; the Mediterranean up to the Middle East; the waters across the Southern Pacific, Atlantic, and Indian Oceans; and a particularly strong correlation in the waters south-east of the Cape of Good Hope. Most of these patterns seem to be associated with the differences in sea ice and ocean dynamics present in the stadial and interstadial states.

The regions in the North Pacific and the North Atlantic (above 40$^\circ$N) that show significant variability are in the areas encompassed by the sea ice limits during the simulated stadial and interstadial periods; that is, the area covered by sea ice during the stadial but not during the interstadial. The belt that can be seen in the southern ocean also largely coincides with the limit of the Antarctic sea ice edge. The areas on the East coast of South America and the south-east of the Cape of Good Hope seem to be a response to the ocean dynamics. Meanwhile, the trail that extends through the Mediterranean Sea all the way to the Middle East may be due to heat transport from the atmosphere since similar tongues can be seen in the temperature anomaly maps of other models that underwent a hosing experiment in LGM conditions \cite{kageyama2013climatic}.

\begin{figure}
    \centering
    \includegraphics[width=0.8\linewidth]{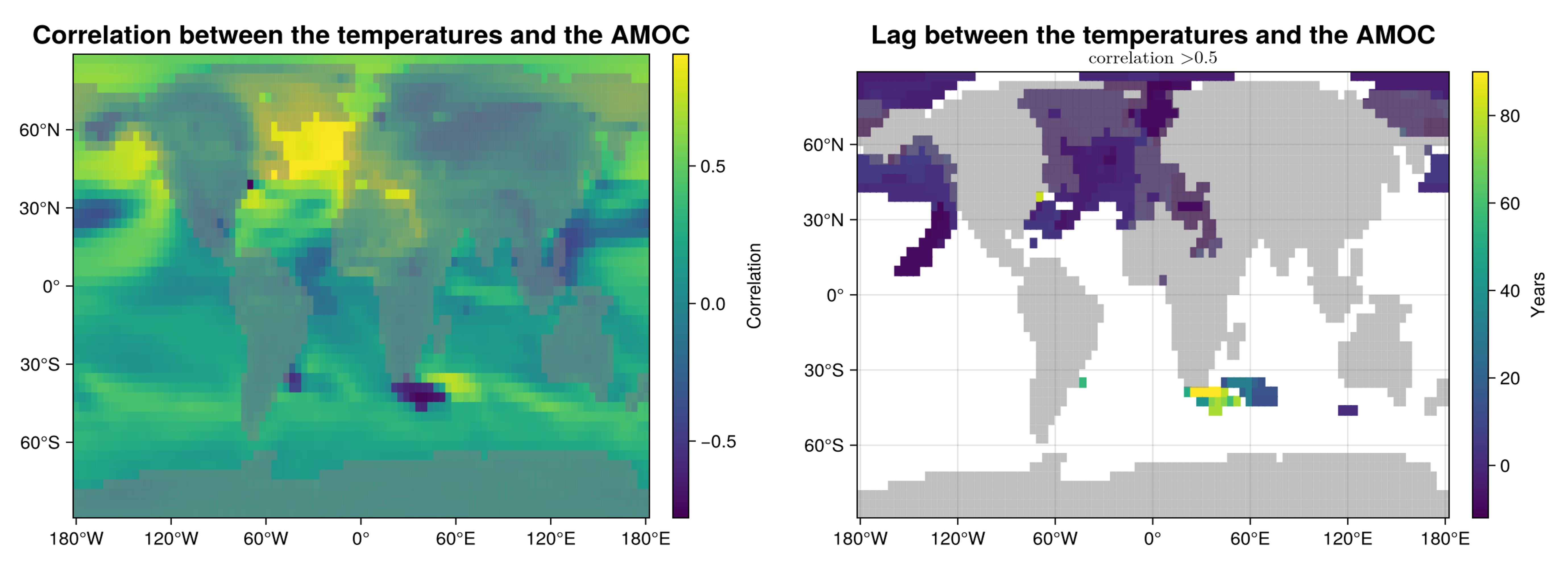}
    \caption{The left panel shows a heat map of the maximum correlation of each data series of annual minima with the AMOC strength, with a lag between $-200$ and $200$ years. The right panel shows a heat map of the lag in years that maximises correlation. Positive lag means that the temperatures lag the AMOC. It is only shown for those data series that show significant correlation ($\geq0.5$).}
    \label{fig:lags_world}
\end{figure}

An important distinction to make is that in this simulation, sometimes the temperatures lag behind the AMOC changes, and sometimes the AMOC lags behind the temperatures. The mechanism of AMOC collapse in this simulation is given by an oscillation of the temperatures, the ocean dynamics, and the sea ice; see \cite{vettoretti2022atmospheric} for a detailed description. It is also worth noting that feedbacks are present between the temperatures, the sea ice, and the AMOC strength, which makes the distinction between cause and effect diffuse. A discussion of different mechanisms for D-O oscillations and whether changes in AMOC are a trigger or a consequence of the oscillations can be found in \cite{boers2018ocean}. Figure~\ref{fig:lags_world} shows the lag of the temperatures with respect to the AMOC strength given by a cross-correlation analysis (positive lag meaning that the temperature changes after the AMOC). For most of the data series in the Arctic and the north Atlantic, the AMOC lags the annual minimum temperature. This is particularly intense in the North Atlantic, where changes in atmospheric temperature and sea ice are strongly associated with the strength of the AMOC. The temperatures lag the AMOC in other parts of the world, notably in the North Pacific and in the south-east of the Cape of Good Hope, where correlations are still strong ($\geq 0.5$). 

\subsection{Trends on $\mu$}

Trends in $\mu$ signify a translation of the distribution of minima. The top panel of Figure~\ref{fig:planetplot} shows the magnitude of $\mu_1$, which would be akin to the sensitivity of $\mu$ to changes in AMOC strength if there were a causal relationship. 

There are significant positive correlations between the AMOC strength and the $\mu$ parameter almost everywhere around the world. It is remarkably stronger in the areas stated before: the North Atlantic stretching to the Middle East, the North Pacific, and across the southern Pacific, Atlantic, and Indian oceans.

There is a negative correlation as well in other areas of the Atlantic, such as the east coast of North America and the east coast of south America near Río de la Plata, in some inland areas such as Alaska, Central Asia, and Siberia, and in the southern seas to the south-east of the Cape of Good Hope.

\subsection{Trends on $\sigma$}

Trends in $\sigma$ signify a change in the spread of the distribution of minima, thus a change in the variability of the observed extremes. The middle panel of Figure~\ref{fig:planetplot} shows the magnitude of $\sigma_1$, which would be akin to the sensitivity of $\sigma$ to changes in AMOC strength, if there were a causal relationship. 

There is an interesting pattern that repeats in several places near the sea ice edge. Areas with strong positive correlation are very close to areas with substantial negative correlation. This happens most notably in the North Atlantic and the North Pacific regions, and on a smaller scale in the Southern seas. This positive correlation is sometimes a bias of the pattern detection, produced by the high variability introduced during the transition between states; see Appendix~\ref{ap:linear_model}. In general, there are significant positive correlations in the North Atlantic stretching to the Middle East, the North Pacific, and the Southern Seas. Strong negative correlation can be observed near the sea ice margins of the interstadial due to its inter-annual variability (south of the north Atlantic and north Pacific areas, and north of the Antarctic ocean) and extending through the Mediterranean to the Middle East. Negative correlation can also be observed in Greenland. This is consistent with the proxy record for temperatures, which shows greater variability during the stadial periods \cite{brashear2025shifts,ditlevsen2002fast}.

\subsection{Trends on $\xi$}

Trends in $\xi$ signify a change in the tails of the distribution of minima, thus a change in the decay of the probability of the extremes with its intensity. The bottom panel of Figure~\ref{fig:planetplot} shows the magnitude of $\xi_1$, which would be akin to the sensitivity of $\xi$ to changes in AMOC strength if there were a causal relationship. 

In most of the world, there are no significant changes in $\xi$ against the AMOC strength. This is partially due to the method for selecting models, which biases towards simpler models. One example of such bias is the lack of signal in the middle of the North Atlantic, where other parameters have the strongest signal. See Appendix~\ref{ap:linear_model} for more detail. 

A positive correlation can be observed in the North Pacific, around the North Atlantic, extending through the Mediterranean to the Middle East, and in some areas around the southern Ocean sea ice edge.

A negative correlation can be observed below the areas of positive correlation in the North Atlantic and North Pacific, and in the spot located south-west of the Cape of Good Hope. 

\section{Spatial Patterns of Variability}

\begin{figure}
    \centering
    \includegraphics[width=\linewidth]{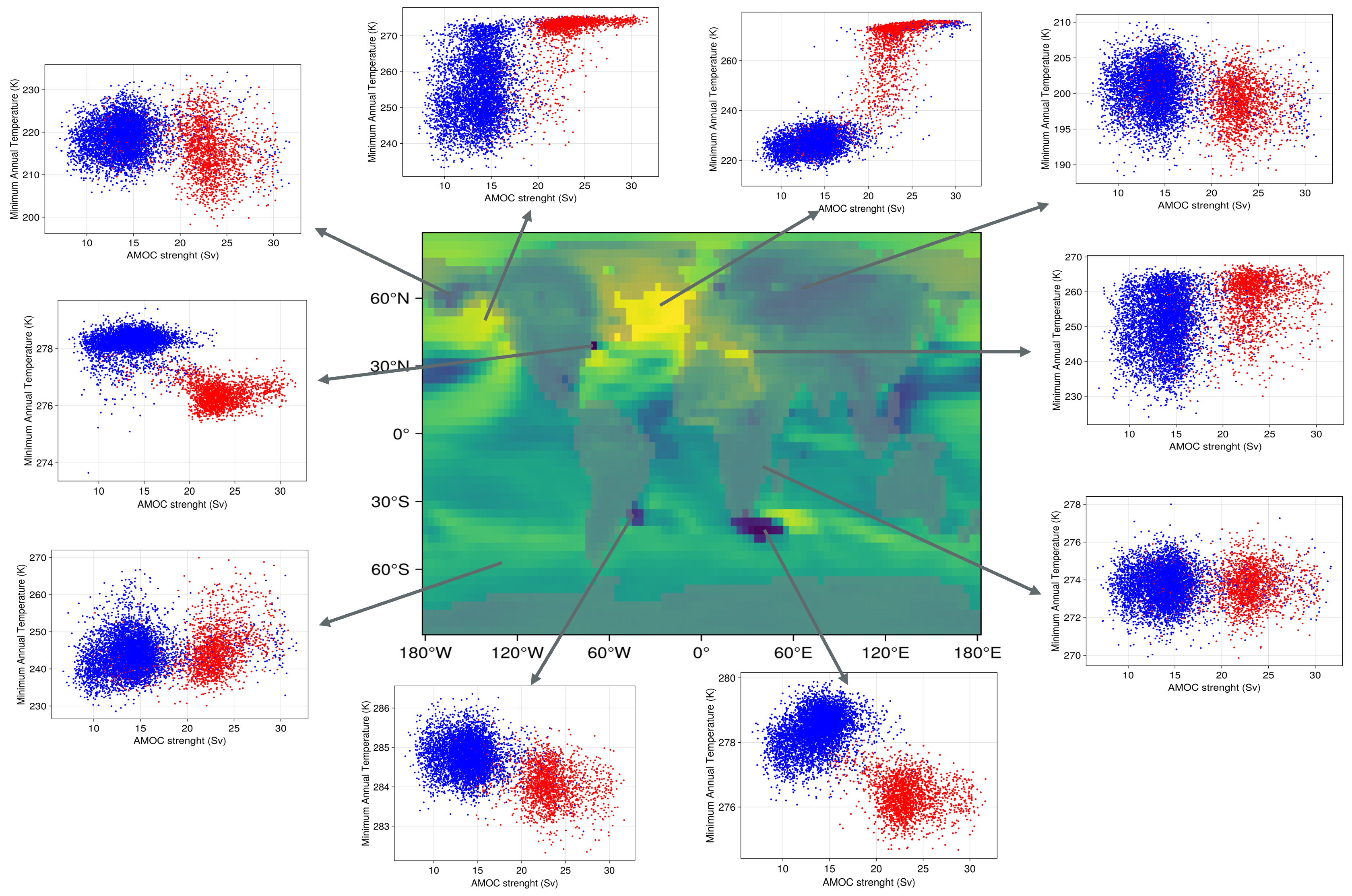}
    \caption{This figure shows different behaviours of the annual minima in the different regions highlighted before. The central panel is the same as the left panel of Figure~\ref{fig:lags_world} to show approximately the locations where the data series are taken.}
    \label{fig:global_areas}
\end{figure}

In this section, we take a more detailed look at the regions in the world where significant patterns are found. Figure~\ref{fig:global_areas} shows the minimum annual temperature plotted against the strength of the AMOC in some exemplary locations, highlighted by the linear model.

\subsection{Non-parametric model}

One of the most obvious learnings from Figure~\ref{fig:global_areas} is that different areas of the world exhibit very different behaviours. The linear model (Eq.\ref{eq:linear_model}), while useful for trend detection, does not necessarily capture the dependency of the parameters on the AMOC strength. Thus, a non-linear analysis of the parameters requires individual treatment for the regions. For that purpose, we compute another fit to the parameters without prescribing a particular relationship between them. The minima are sorted according to the AMOC strength value, and then the fit to a stationary GEV is computed over a rolling window on the AMOC strength. This is a non-parametric fit (as opposed to the linear model) for the functions $\mu(\mathrm{AMOC}(t))$, $\sigma(\mathrm{AMOC}(t))$ and $\xi(\mathrm{AMOC}(t))$, which allows us to observe a more accurate dependency of the parameters as a function of the AMOC strength beyond the linear model. In the following plots (Figs.\ref{fig:Region_north_atlantic}-\ref{fig:Region_south_pacific_and_america}), we have plotted the parameters $\mu(\mathrm{AMOC}(t))$ in the top, $\sigma(\mathrm{AMOC}(t))$ in the middle, and $\xi(\mathrm{AMOC}(t))$ in the bottom for selected locations. The original data points are plotted in black in the background to help with the visualization of the underlying distribution being modelled, but they bear no relationship to the axis. 

\subsubsection{North Atlantic}

\begin{figure}
    \centering
    \includegraphics[width=0.5\linewidth]{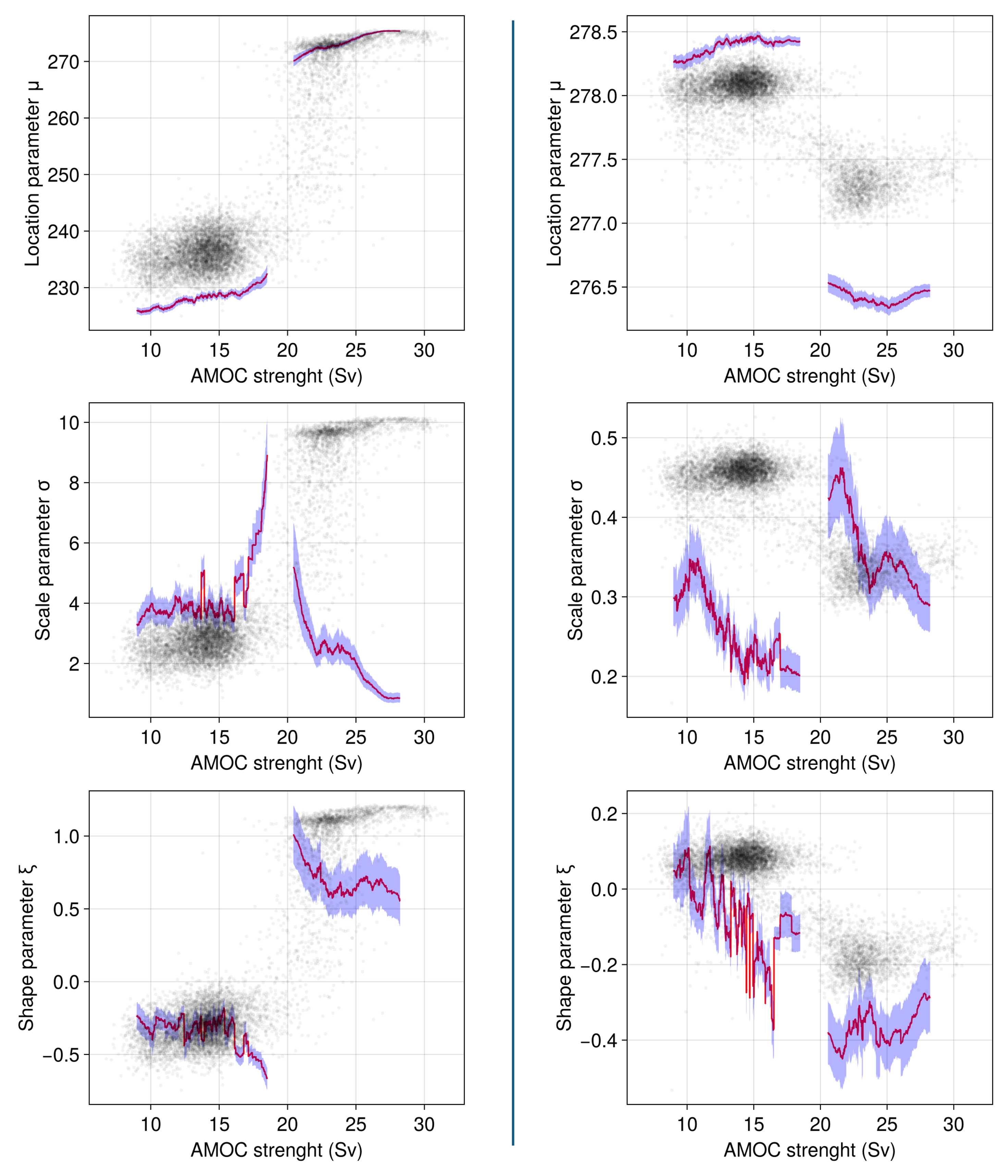}
    \caption{This figure shows how the parameters change against the AMOC strength. The left panel corresponds to the location 30$^\circ$W 57$^\circ$N, approximately the middle of the north Atlantic, and the right panel to the location 71$^\circ$W 39$^\circ$N, the inversely correlated spot in the East Coast of North America.}
    \label{fig:Region_north_atlantic}
\end{figure}

Figure~\ref{fig:Region_north_atlantic} shows how the parameters change in two selected locations. On the left, a data series belonging to the middle of the North Atlantic is presented. This region is subject to the largest temperature changes. Most of this region is covered by sea ice during the stadial period and is ice-free during the interstadial period. This fact explains most of the variability observed: high minimum temperatures during the interstadial and low minimum temperatures during the stadial, resulting in a high $\mu_1$ as seen in the first panel of Figure~\ref{fig:planetplot}. 
It can also be seen that the scale decreases in the interstadial compared to the stadial, which can be interpreted as the minima having less variability due to the absence of sea-ice. A negatively correlated trend can also be seen where the stadial edge of the sea ice is. The variability will be greater during the stadial in these coordinates because some years there will be sea ice and some years there will not, which will have a strong effect.  

On the right, this corresponds to the spot on the East coast of North America that is inversely correlated with AMOC strength. In this spot, the temperatures significantly lag behind the changes in AMOC strength, and the temperatures do not go below freezing point. Some other models show a similar warming of this region in a hosing experiment \cite{kageyama2013climatic}. Interestingly, the parameters seem to change in an almost symmetrical fashion with respect to the other location chosen.

\subsubsection{North Pacific}

\begin{figure}
    \centering
    \includegraphics[width=0.5\linewidth]{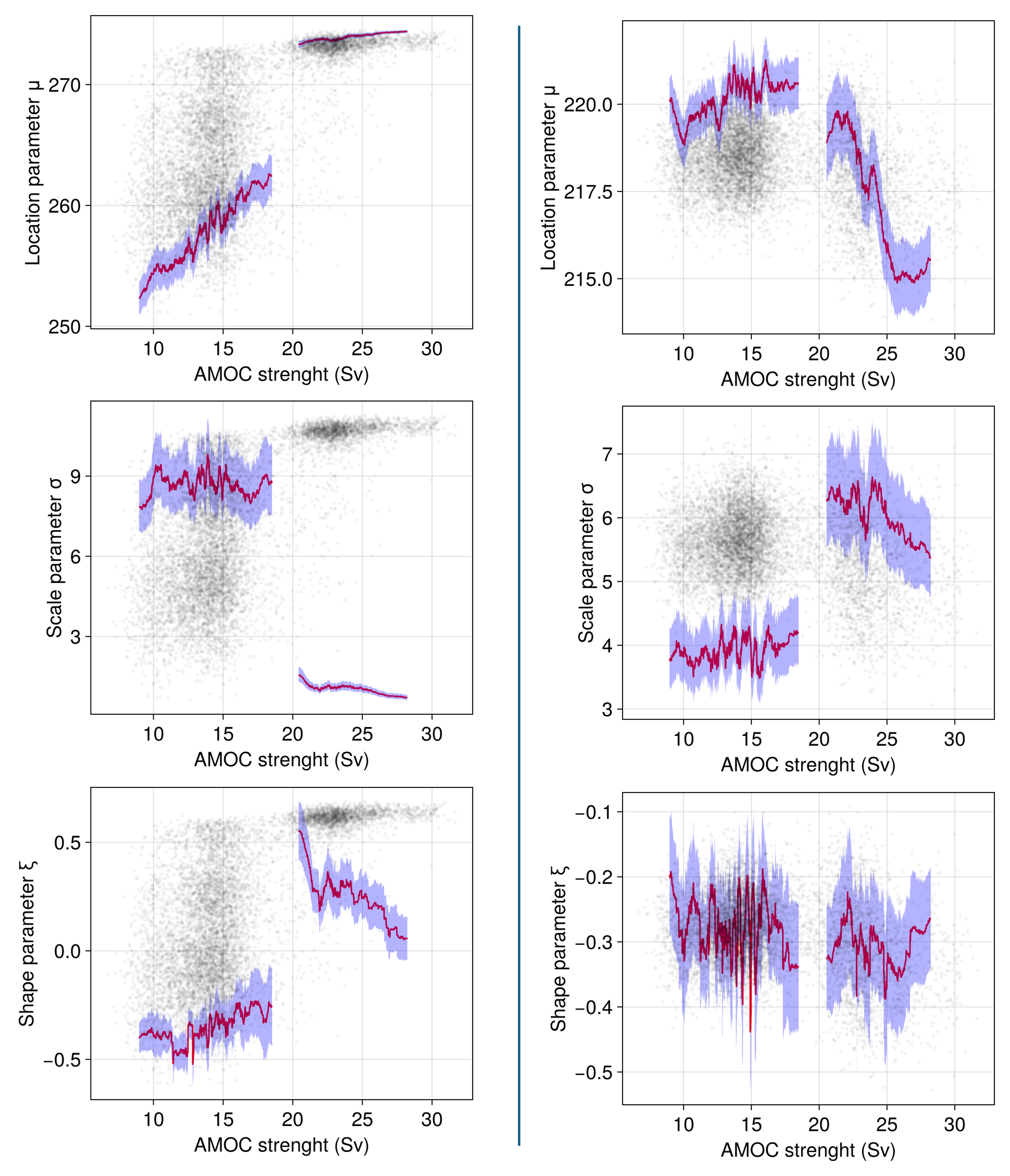}
    \caption{The left panel corresponds to the location 135$^\circ$W 46$^\circ$N, in the Pacific coast of north America, and the right panel to the location 154$^\circ$W 65$^\circ$N, in inland Alaska.}
    \label{fig:Region_north_pacific}
\end{figure}

Figure~\ref{fig:Region_north_pacific} shows how the parameters change in two locations in Alaska and the Alaskan Gulf. In the left, the waters of the Alaskan Gulf show the same phenomena as the North Atlantic region regarding the sea ice. It is worth noting that the changes in temperature lag behind the AMOC strength. The sea ice expansion happens after the AMOC slowdown and induces the teleconnection that can be observed with the North Atlantic. The proxy record of lake sediments from \cite{anderson2026shifting} shows this teleconnection mechanism between the temperatures in Alaska and the AMOC in the North Atlantic during the end of the LGM. On the right, the temperatures in Alaska are negatively correlated with the AMOC strength inland. The location and the scale seem to change significantly, while the shape parameter is approximately constant.

\subsubsection{Middle East and Siberia}
\begin{figure}
    \centering
    \includegraphics[width=0.5\linewidth]{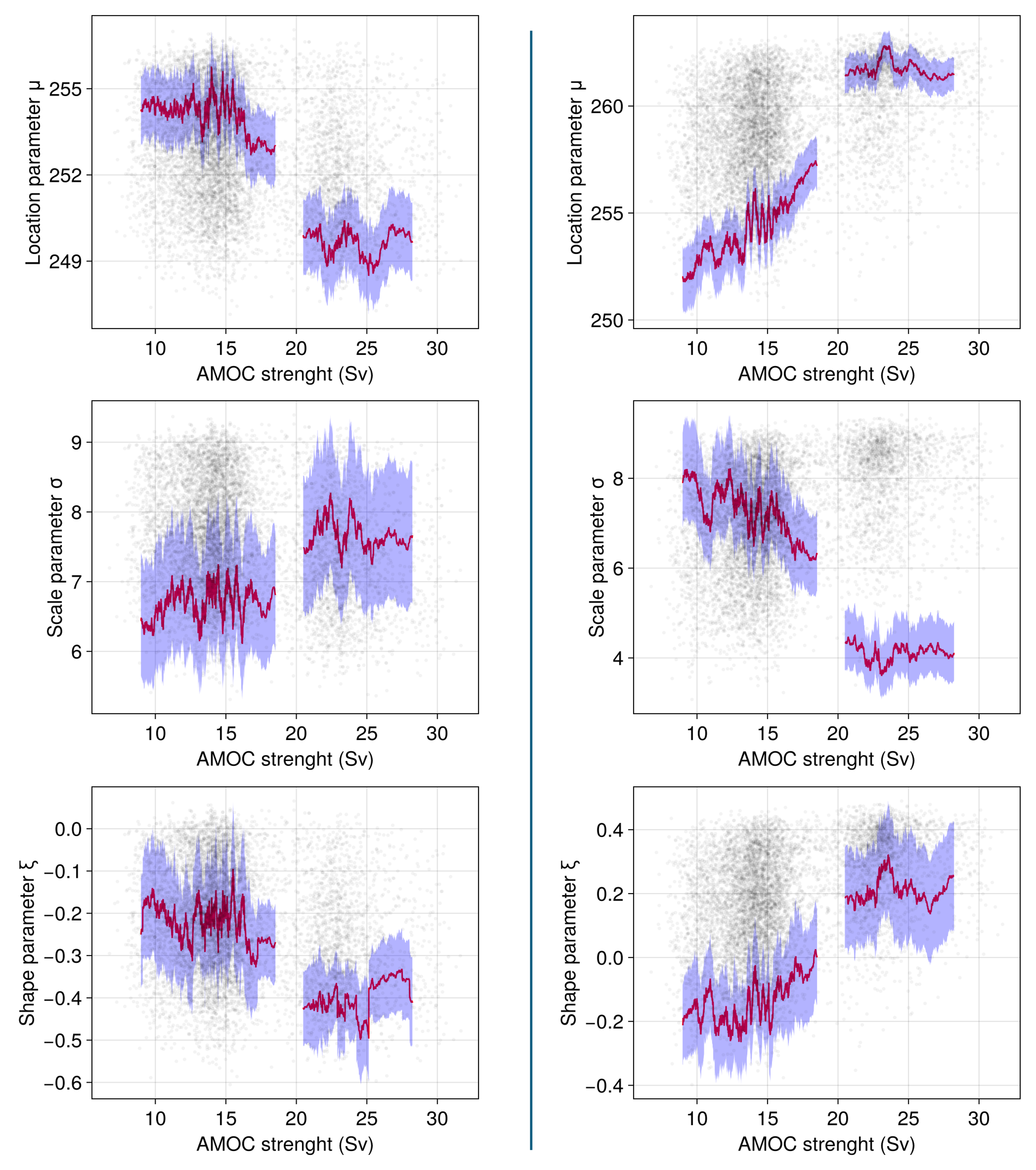}
    \caption{The left panel corresponds to the location 71$^\circ$E 62$^\circ$N, in inland Siberia, and the right panel to the location 37$^\circ$E 35$^\circ$N, in the Middle East.}
    \label{fig:Region_Mediterranean_and_siberia}
\end{figure}

In the left of Figure~\ref{fig:Region_Mediterranean_and_siberia}, in Siberia, we also see a negative correlation, as in Alaska, although the change only seems to significantly affect the location parameter. In the right, the Middle East region shows a similar correlation to the North Atlantic region, although it is less intense. The three parameters seem to change in a significant manner. An interesting dynamical feature that we see in this location is that the minima seem to be approximately stationary against the AMOC during the interstadial but non-stationary during the stadial. 

\subsubsection{South of Africa}

\begin{figure}
    \centering
    \includegraphics[width=0.5\linewidth]{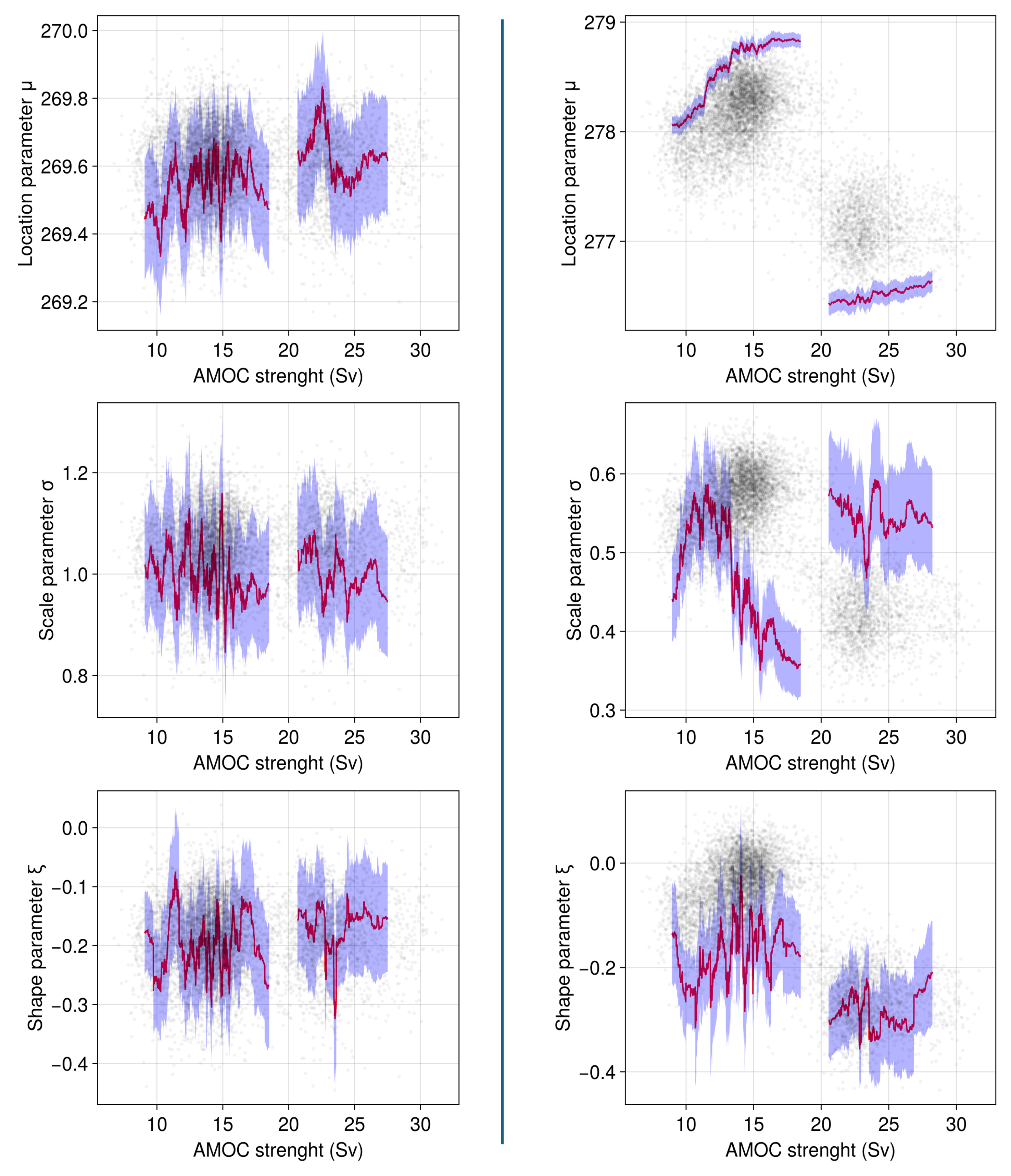}
    \caption{The left panel corresponds to the location 34$^\circ$E 17$^\circ$S, in inland Africa in front of the island of Madagascar, and the right panel to the location 34$^\circ$E 42$^\circ$S, in the blob off the coast of Cape of Good Hope.}
    \label{fig:Region_South_of_Africa}
\end{figure}

Large parts of Central and Southern Africa show stationarity; the annual minima seem to be statistically the same in the stadial and interstadial states. The left panels of Figure~\ref{fig:Region_South_of_Africa} show an example of such a climate. In the right panels, a location in the sea off the Cape of Good Hope is shown. In this location and the locations around it, the temperatures are strongly inversely correlated. The right panel of the figure shows how the parameters vary there, and overall it seems to exhibit the opposite pattern compared to the North Atlantic in all parameters.

\subsubsection{South Pacific and South America}

\begin{figure}
    \centering
    \includegraphics[width=0.5\linewidth]{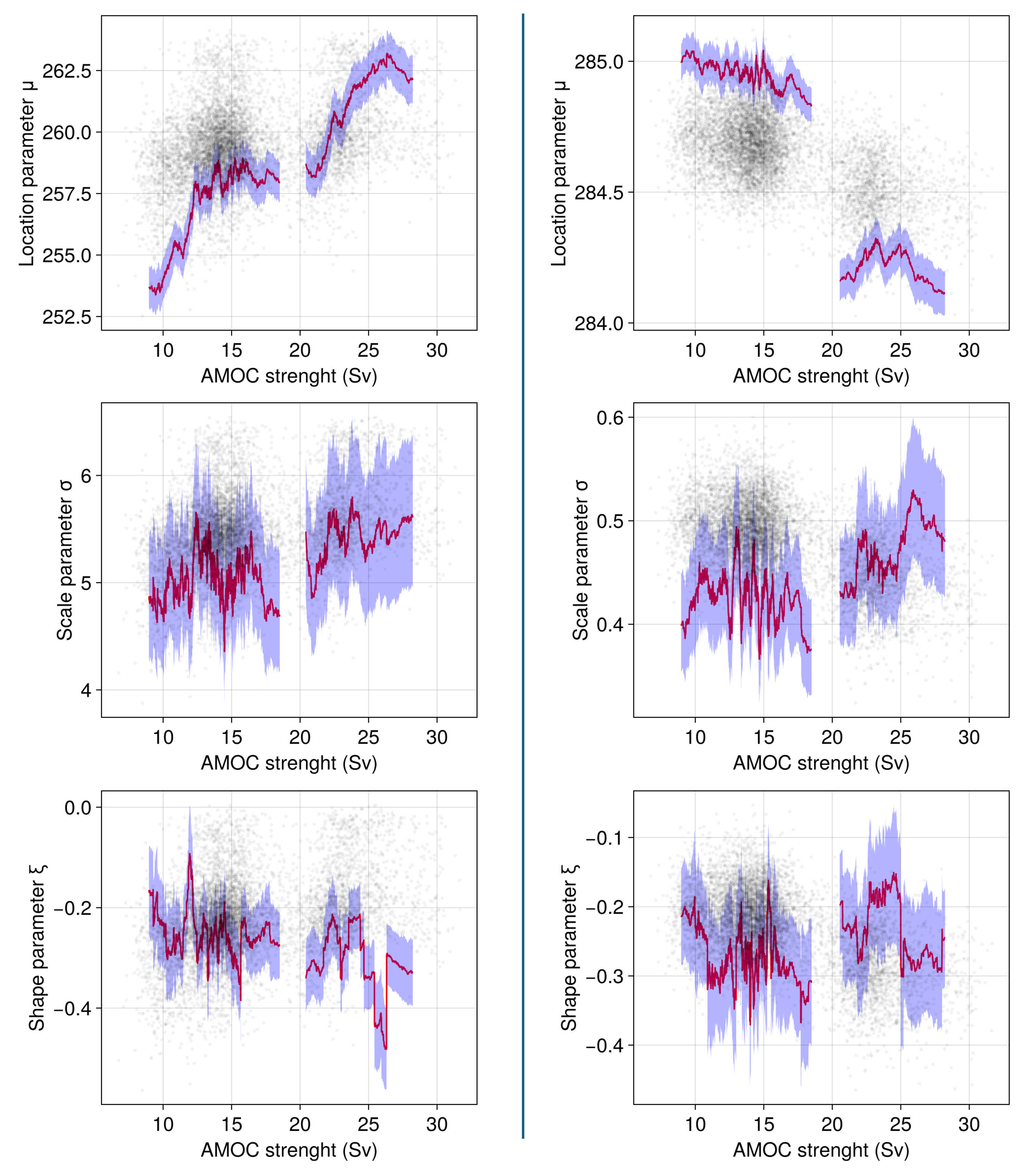}
    \caption{The left panel corresponds to the location 131$^\circ$W 57$^\circ$S, in the southern Pacific, and the right panel to the location 45$^\circ$W 35$^\circ$S, off the coast of Río de la Plata.}
    \label{fig:Region_south_pacific_and_america}
\end{figure}

The left panels of Figure~\ref{fig:Region_south_pacific_and_america} correspond to a location in the South Pacific. This location is near the Antarctic sea ice edge, which does not change as strongly between stadial and interstadial in our model. The changes in the parameters are different from those observed in the locations near the sea ice edge of the Northern Hemisphere, with only the location changing significantly and an apparent increasing trend in the scale parameter.

The right panels correspond to the inversely correlated spot that can be observed in Figure~\ref{fig:lags_world} near Rio de la Plata, on the Atlantic coast of South America. The change is most notable in the location parameter, contrasting with the other inversely correlated spots in the sea in Figs.~\ref{fig:Region_north_atlantic} and \ref{fig:Region_South_of_Africa}.

\subsection{Analysis}

Many different non-linear trends are realized by the data on the model. In many cases, especially those showing strong correlation as per Figure~\ref{fig:lags_world}, more than one parameter shows a trend. In many cases, trends within a state (stadial/interstadial) are linear or approximately constant, with a strong step or peak in the middle, corresponding to the values where the dynamics transition between states. The tail parameter experiences significant changes in the areas where strong variability between the states is present. All combinations of stationarity and non-stationarity are realized; see Appendix~\ref{ap:linear_model} for a detailed discussion.

\section{Physical drivers of variability}

A change in the heat budget of a region is a sufficient physical mechanism for changes in $\mu$, since it shifts the distribution of temperatures. However, different mechanisms are needed to explain changes in $\sigma$ or $\xi$ \cite{huang2016estimating}. Indeed, in Figure~\ref{fig:planetplot}, one can see that changes in $\mu$ are widespread, while changes in $\sigma$ and $\xi$ seem to follow patterns that we can associate with the physical processes of the oceans and the cryosphere. 

\subsection{Sea Ice}

As discussed above, the sea ice is one of the major drivers of variability for the minima. During the stadial periods, the Arctic sea ice goes as far south as $40^\circ$N in the Atlantic and the Pacific, while during the interstadial, the North-Eastern parts of the oceans are ice-free. This pattern can be seen in all panels of Figure~\ref{fig:planetplot} and in the correlations of Figure~\ref{fig:lags_world}. 

By comparing the distributions that arise in the North Atlantic and Pacific when sea ice is present in the stadial and absent in the interstadial, we conclude the following: The presence of sea ice shifts the distributions of minima to colder extremes (decrease in $\mu$), increases the variability of the extremes (increase in $\sigma$), and also skews the distributions of extremes towards less frequent colder extremes (decrease in $\xi$) generally. Near the sea ice edge, inter-annual variability at some coordinates implies that some years there is ice cover and some years there is not. This induces extreme variability that translates into a substantial increase in $\sigma$ and/or $\xi$.

\subsubsection{Fat tails}

Here we will refer to fat tailed distributions as those GEV distributions for which the variance does not converge. When $\xi\geq 0$ the data exhibit strong variability, and observables of the form $\frac{1}{n}\sum(x_i-\Bar{x})^{1/\xi}$ do not converge as $n\to\infty$. This includes the moments of the distribution, including the variance when $\xi\geq 1/2$ and the mean when $\xi\geq 1$ for which $\Bar{x}= \frac{1}{n}\sum x_i$ does not converge either. Thus, fat tails are characterized by a tail parameter bigger than $\xi\geq 1/2$, and they can be interpreted as a signature of very strong variability in the data and strong skewness. 

Although fat tailed distributions are rather extreme in this sense, there is some evidence that they are an adequate statistical model for some geophysical processes. Distributions without a variance have been found in the high frequency variability of the Greenland ice core record \cite{ditlevsen1999observation}, and can appear in the distributions of hydrological variables as well \cite{katz2012statistical}. 

\begin{figure}
    \centering
    \includegraphics[width=\linewidth]{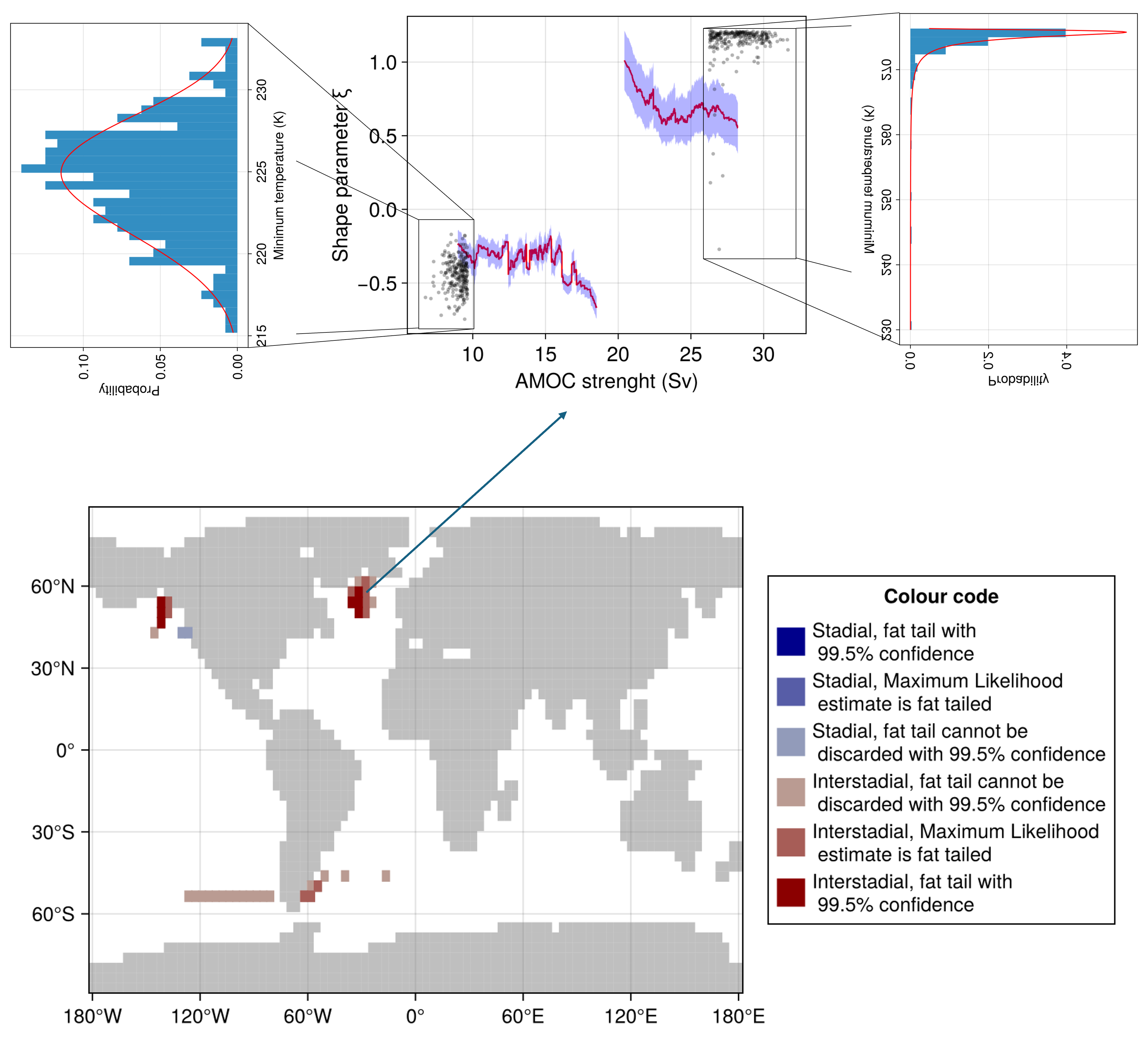
}
    \caption{Visualisation of a fit that produces fat tails in the interstadial state. The middle plot shows the non-parametric fit of the shape parameter, and the black dots correspond to the 250 minimum temperatures associated with lowest AMOC strength (stadial) and the 250 associated with the highest AMOC strength (interstadial). The left and right panels are the histograms together with the GEV fit that produces the shape parameter estimate. The right panel is mirrored to facilitate comparison. The data series used corresponds to 30$^\circ$W 57$^\circ$N. The bottom plot shows the location of the fat tails. The stronger the colour, the stronger the confidence that a fat tail exists. Blue corresponds to fat tails during the stadial and red during the interstadial.}
    \label{fig:fat_tails}
\end{figure}

Figure~\ref{fig:fat_tails} shows where fat tailed GEV distributions ($\xi\geq 1/2$) appear to be the correct model for the minimum annual temperatures in our model. To avoid the instability induced by the transition between interstadial and stadial, only the point in the extremes of the AMOC strength range are considered. The colour code of the figure represents the confidence that a fat tail exists, and whether it exists in the stadial or the interstadial state. Same as before, blue means that it takes place in the stadial and red in the interstadial. The faint colour means that there are fat tails within the 99$\%$ confidence interval of the tail parameter, medium intensity means that the maximum likelihood estimate for the tail parameter is $1/2$ or above, and the intense colour means that all the 99$\%$ confidence interval lies above $1/2$. The fat tailed distributions shown in Figure~\ref{fig:fat_tails} are located along the borders of the sea ice, both for the stadial and the interstadial states. It is worth noting that in the location chosen for the top panel, while the distribution of minima during the stadial is approximately Gaussian, the distribution corresponding to the interstadial is very skewed and strongly non-Gaussian.

\subsection{Ocean dynamics}

The increase or decrease of the AMOC strength is part of a reorganisation of the currents not only in the North Atlantic but all over the globe. Particularly, the weakening and strengthening of the AMOC is preluded by changes in the North Atlantic deep water formation and the Antarctic Bottom Water \cite{vettoretti2022atmospheric}. 

At least two big spots located in the sea show significant negative correlation ($\leq-0.5$) and lag the AMOC strength 60 to 80 years. Moreover, both spots are located in places where main current systems converge and change direction \cite{kampf2016upwelling}. One of such spots is the blob South East of Cape of Good Hope. It is located where the Agulhas current travelling south-west along the African coast meets the strong eastward Antarctic Circumpolar Current and changes direction (Agulhas Current Retroflexion). The other spot is located off the coast of Río de La Plata. In this spot the western boundary Brazil current meets the northern bound Falkland/Malvinas current and changes direction (Brazil-Falkland/Malvinas Confluence). Changes in the strength of the AMOC are related to changes in the strengths of these currents \cite{pontes2024weakening}, particularly an strengthening of the Antarctic Circumpolar Current. This is also observed in our model, along with 
persistent anomalies in the Sea Surface Temperature (SST) and the zonal and meridional velocity respectively, again strongly correlated with the AMOC strength. It is interesting to note that in the area of the Agulhas current retroflexion we observe opposite trends as in the North Atlantic in all parameters, decrease of $\mu$, increase of $\sigma$ and decrease of $\xi$ when AMOC strength increases, whereas in the Brazil-Falkland/Malvinas Confluence $\sigma$ and $\xi$ are approximately stationary while the location parameter $\mu$ shows a similar decreasing trend.

There are several sediment cores available that cover the Southern Atlantic and the Agulhas region \cite{gersonde2003last}. In the cores there are fossils of diatoms and chemical compounds such as alkenones that can be used to create an approximate reconstruction of the SST during the LGM. Further detail about the proxies discussed and their intricacies can be found in References \cite{gersonde2003last,verma2023variation,romero2015high}. In the cores studied in these papers there is no direct evidence of strong temperature anomaly inversely correlated with the AMOC in the same areas that our model does. Moreover the papers focus in comparing the glacial and interglacial periods while our focus is in comparing stadial and interstadial states. However, there are multiple lines of evidence of the shifting of the currents affecting significantly the temperatures in these regions. In the cores situated along the coast in the South-west of Africa, \cite{verma2023variation} finds maxima of SST during glacial periods, which the authors attribute to a change in the intensity of the Agulhas leakage, waters from the warm Indian Ocean that pass to the South Atlantic. In \cite{romero2015high} the reconstruction of the SST temperatures in the region South East of Cape of Good Hope is correlated with the data form the Vostok ice core. A parallelism that we observe is that the Vostok ice core is correlated with NGRIP ice core through the bipolar see-saw model from Ref.\cite{stocker2003minimum}. Similarly we observe strong correlation between the anomalies in temperature in this location and same see-saw model applied to the North Atlantic temperature anomalies temperatures from our model. In this same region, the cores on \cite{gersonde2003last} signal less cooling than in the rest of the South Atlantic during the glacial periods. In this paper also a core near Rio de la Plata is analysed although it is from a location further south of the spot we see on our model. This core shows anomalously high cooling during the glacial periods, which the authors link to less southward penetration of the warm Brazil Current.

\section{Extrapolation and Limitations}
\subsection{Mechanisms of AMOC decline}
One of the overarching ideas of this research is to use the data of the past to inform future projections. The possibility of a AMOC weakening has become a realistic future scenario \cite{portmann2026observational,van2025european}. 
Different climate models show an AMOC slowdown or collapse when forced by an increase of CO$_2$ concentration \cite{portmann2026observational}, hosing with freshwater in the North Atlantic region in LGM conditions, PI conditions or various Shared Socio-economic Pathways (SSP) scenarios \cite{kageyama2013climatic,van2025european} or even just by internal variability of the system in present day climate conditions \cite{cini2024simulating}. There are thus different mechanisms for AMOC slowdown and collapse and different climate states that can be reached as a consequence. A warm future with AMOC collapse would happen in very different conditions than the transition between interstadial and stadial states that we observe in our data.
However, the parameters of the GEV distributions seems to follow almost linear trends as a function of the AMOC strength in many of the locations in the stadial and interstadial state, and that could be a feature that stays stable across some models and some climate states. 


The simulation we employ is set in a hypothetical LGM conditions. The boundary conditions are LGM but the evolution is constrained by a fixed level of atmospheric CO$_2$ and a fixed ice sheets \cite{vettoretti2022atmospheric}. 
The spontaneous D-O oscillations occur in this setting in a window of CO$_2$ between 170 and 240 ppm \cite{vettoretti2022atmospheric}. For these reasons our current dataset does not immediately extend to higher CO$_2$ scenarios. Also the sea level and other key climate variables are fixed to LGM levels, which makes the simulation rather unphysical, and difficult to extend to other climate states. Remarkably, even in this context, the stadial and interstadial states remain fairy consistent, albeit with some degrees of warming of difference.
\begin{figure}
    \centering
    \includegraphics[width=0.7\linewidth]{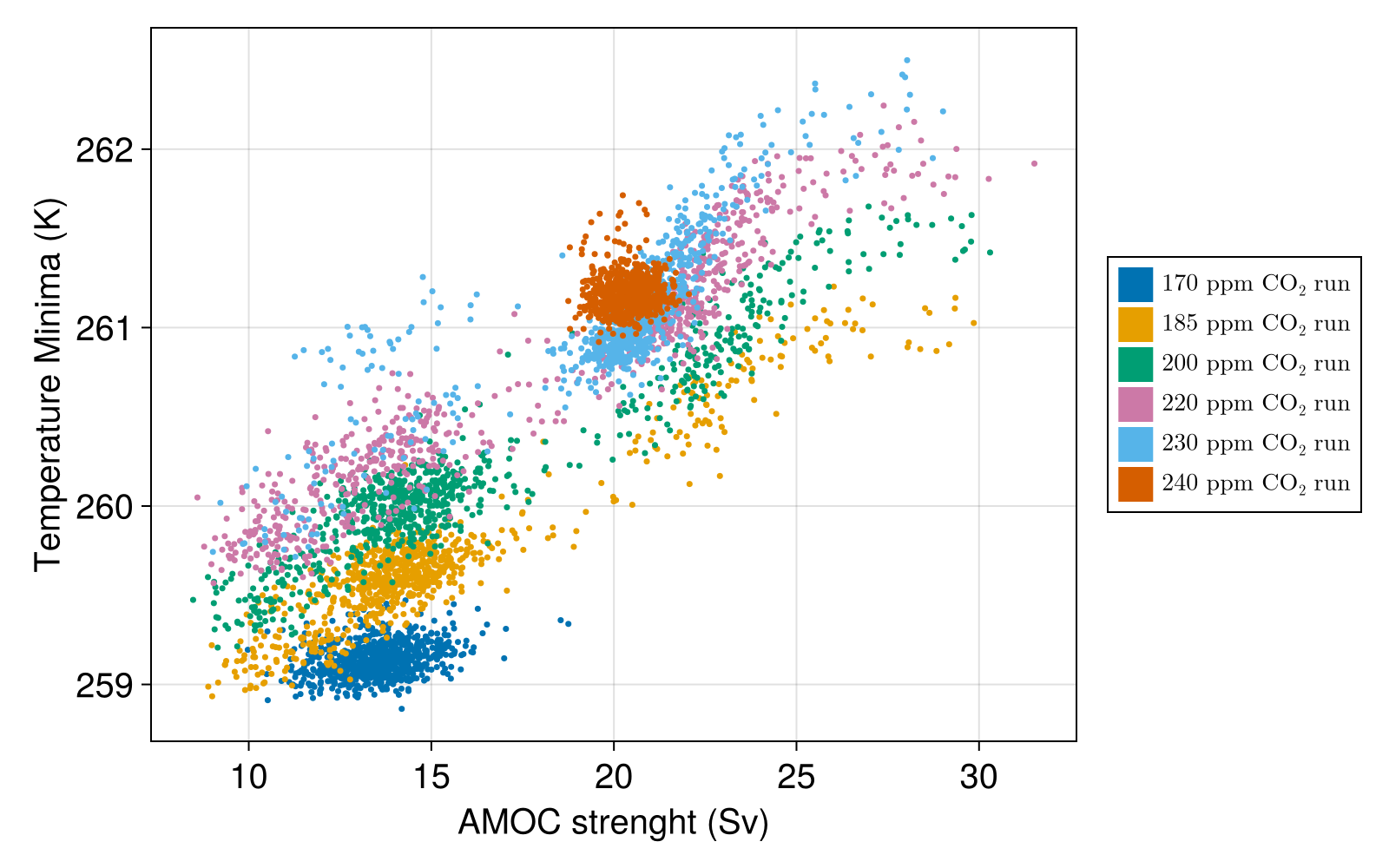}
    \caption{Minimum temperatures averaged spatially across the globe plotted against the AMOC strength for the different runs with different levels of CO$_2$ from \cite{vettoretti2022atmospheric}.}
    \label{fig:different_runs}
\end{figure}
Figure \ref{fig:different_runs} shows how the different concentrations of atmospheric CO$_2$ affect the averaged minimum temperatures. Here we see that the stadial and interstadial states exist for different CO$_2$ concentrations and the temperatures have a seemingly linear dependence on the AMOC within each state. Higher CO$_2$ concentration makes the states slightly warmer, but the overall structure of the states is retained except outside of the CO$_2$ window where D-O oscillations exist. 

In other hosing experiments that study temperatures after an AMOC slowdown in a LGM climate show different temperature anomaly patterns that the switching between stadial and interstadial in our model \cite{kageyama2013climatic}. In all models studied in \cite{kageyama2013climatic} where there is a significant slowdown of the AMOC (including CCSM3) we observe various patterns of cooling in the North Atlantic and heating in the South Atlantic. The magnitude of the changes is much lower that the changes that we observe in the D-O oscillations, but their slowdown of the AMOC through hosing is not a transition between an interstadial state and a stadial. For example, patterns such as the heating in the South Atlantic are present in many models of \cite{kageyama2013climatic}, but not in our simulations. We believe that the associated heating in the South Atlantic is offset by the general cooling of the stadial, which does not happen when hosing. Interestingly hosing experiments have been done with HadCM3 both in 1980 climate and LGM conditions and they result in quite similar anomaly patterns \cite{kageyama2013climatic,jackson2015global}. This suggests than the transitions between stadial and interstadial are fundamentally different that just a collapse of the AMOC in interstadial conditions. In this case, the AMOC slowdown is just one global scale physical process in a larger transition between climate states, rather than being the fundamental driving difference, like in hosing experiments.

For extrapolation to other states, it might be better to use the average temperature that the AMOC strength as covariate. Changes in the stadial-interstadial state and in the atmospheric CO$_2$ are reflected in the average temperature, which could give a more consistent way of dealing with global warming scenarios and/or potentially other stable states of the climate system. However this could also shift the estimations such that a warming and a collapse of the AMOC leave the system in the stadial regime according to the temperatures. A successful extrapolation likely needs to take in account many variables that summarize the individual processes (global and local) that influence a location, if such is possible. In the following section we explore this ideas tentatively.

\subsection{Extrapolation}

To our knowledge only one other study exist in the literature where return levels have been computed for two alternate states of the AMOC \cite{van2025european}. In this study the return levels for 10 year events in a pre-industrial (PI) climate with two different states of the AMOC (ON/OFF), and in a RCP 4.5 scenario also with AMOC (ON/OFF) states is computed. Due to the reasons above here we try to compare their PI (ON/OFF) climates to our interstadial and stadial climates respectively.

\begin{figure}
    \centering
    \includegraphics[width=0.5\linewidth]{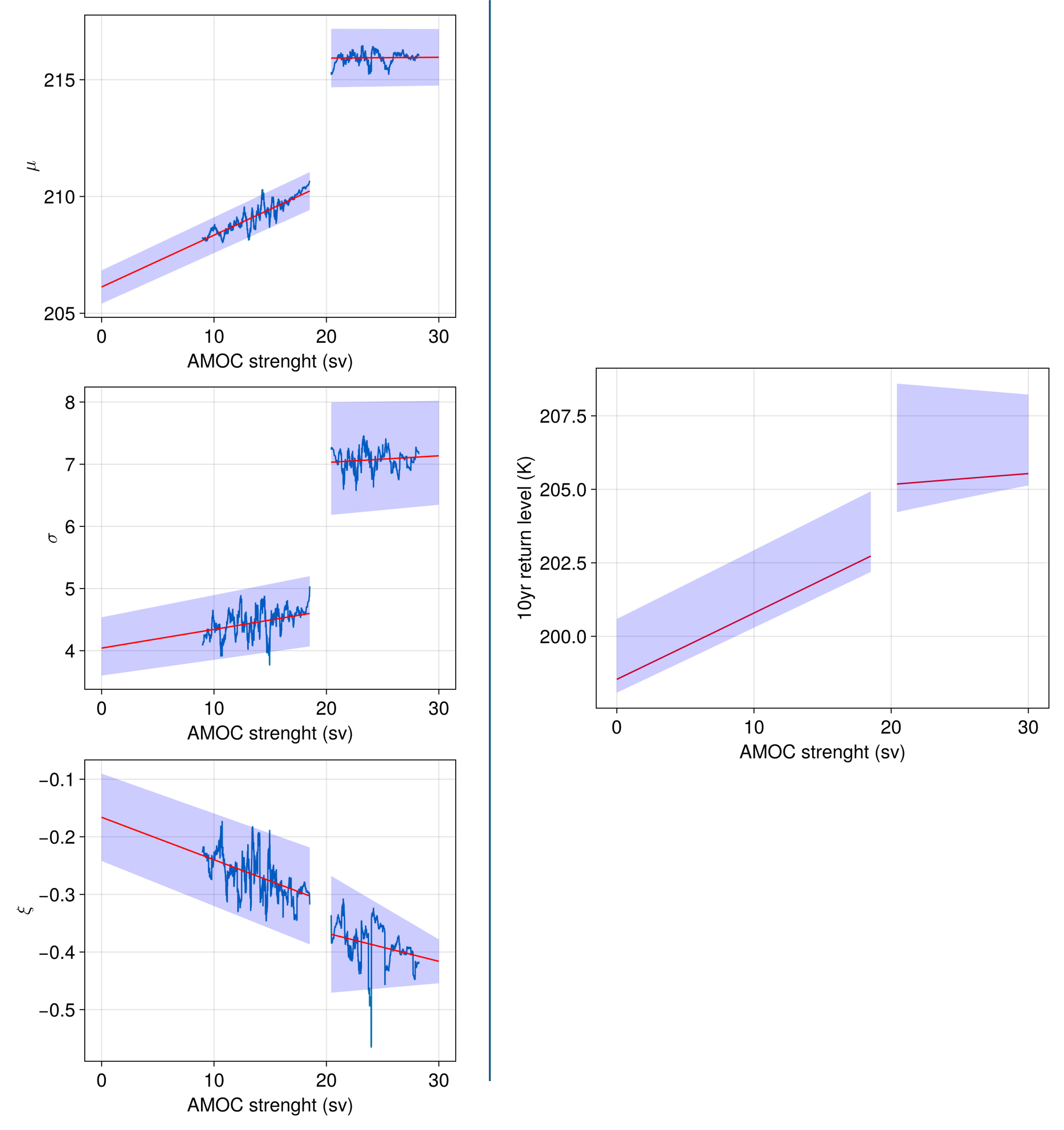}
    \caption{Extrapolation to the approximate location of De Bilt where there is another estimate of return time of cold extremes in their system \cite{van2025european}.}
    \label{fig:extrapolation_De_Bilt}
\end{figure}

Extrapolating in the location chosen in \cite{van2025european} (De Bilt, Netherlands) leads very different results in our system. In \cite{van2025european} the PI AMOC ON state has a 10 year return level of approximately -5 C$^\circ$ at a AMOC strength of approximately $16$ Sv, and the PI AMOC OFF state has a 10 year return level of approximately -35 C$^\circ$ at a AMOC strength of approximately $4$ Sv. Figure~\ref{fig:extrapolation_De_Bilt} shows an attempt to compute the same 10 year return levels for the same approximate location in our system with the same strengths of the AMOC. 

To extend the range of the parameters to other AMOC values, we fit linear trends in the different states. Figure~\ref{fig:planetplot} shows that a linear trends are preferred to a stationary model with high confidence. Once the parameters have been extended to an adequate range we can see that in our system the 10 year return level for 16 Sv of AMOC strength is approximately -71 C$^\circ$ and for 4 Sv of AMOC strength is approximately -74 C$^\circ$. This estimation does not reproduce adequately neither the value of the return levels nor the relative difference between the states ON and OFF in \cite{van2025european}.  

There are many reasons why this extrapolation did not work.
\begin{enumerate}[a)]
    \item The location is near the sea ice edge in the PI AMOC OFF and far away during the PI AMOC ON of \cite{van2025european}, which increases variability between the states. In our system, the location is not near the sea at all.
    \item The location is coastal in the PI climate, but it is not coastal in the LGM climate.
    \item The location is not far from the Fennoscandian Ice sheet in the LGM conditions, which does not exist in the pre-industrial run.
    \item The AMOC displays a different strength range in the different simulations, going from 0 to 18 Sv approximately in \cite{van2025european} and from 10 to 28 Sv in \cite{vettoretti2022atmospheric}. In fact this discrepancy is specially significant for the reason that the strength corresponding to their PI AMOC ON run of approximately 16 Sv corresponds to our stadial state rather than the interstadial, which should be the climate more similar to the pre-industrial.
    \item The CO$_2$ levels are different between our run (200 ppm CO$_2$) and the pre-industrial conditions (280 ppm CO$_2$) in the atmosphere.
\end{enumerate}

While the points $d)$ and $e)$ represent a global physical difference between the simulations, the points $a)$ to $c)$ are related only to the local characteristics. This means that it might be possible to chose an analogue climate that should be more similar to the PI De Bilt. While analogues are usually chosen based on statistics \cite{reuter2023modelling}, there is no guarantee that having a similar climate in the interstadial state will produce a similar climate in the stadial state. 
 For that reason we set dynamical criteria hoping that similar processes in both states will bring about similar extremes in both states. The criteria to consider it analogous are: 
\begin{itemize}
    \item It has to be in the North Atlantic area
    \item It has to be coastal up to model resolution
    \item It has to be near the sea ice edge in the stadial state
    \item It has to be far from the ice sheets
\end{itemize}

The discrepancies arising from the different AMOC strength range and and the different CO$_2$ levels cannot be mitigated directly, although a location further south will be warmer and potentially compensate for the CO$_2$ difference and the AMOC strength could be rescaled to make the ranges and the transition coincide.

\begin{figure}
    \centering
    \includegraphics[width=0.5\linewidth]{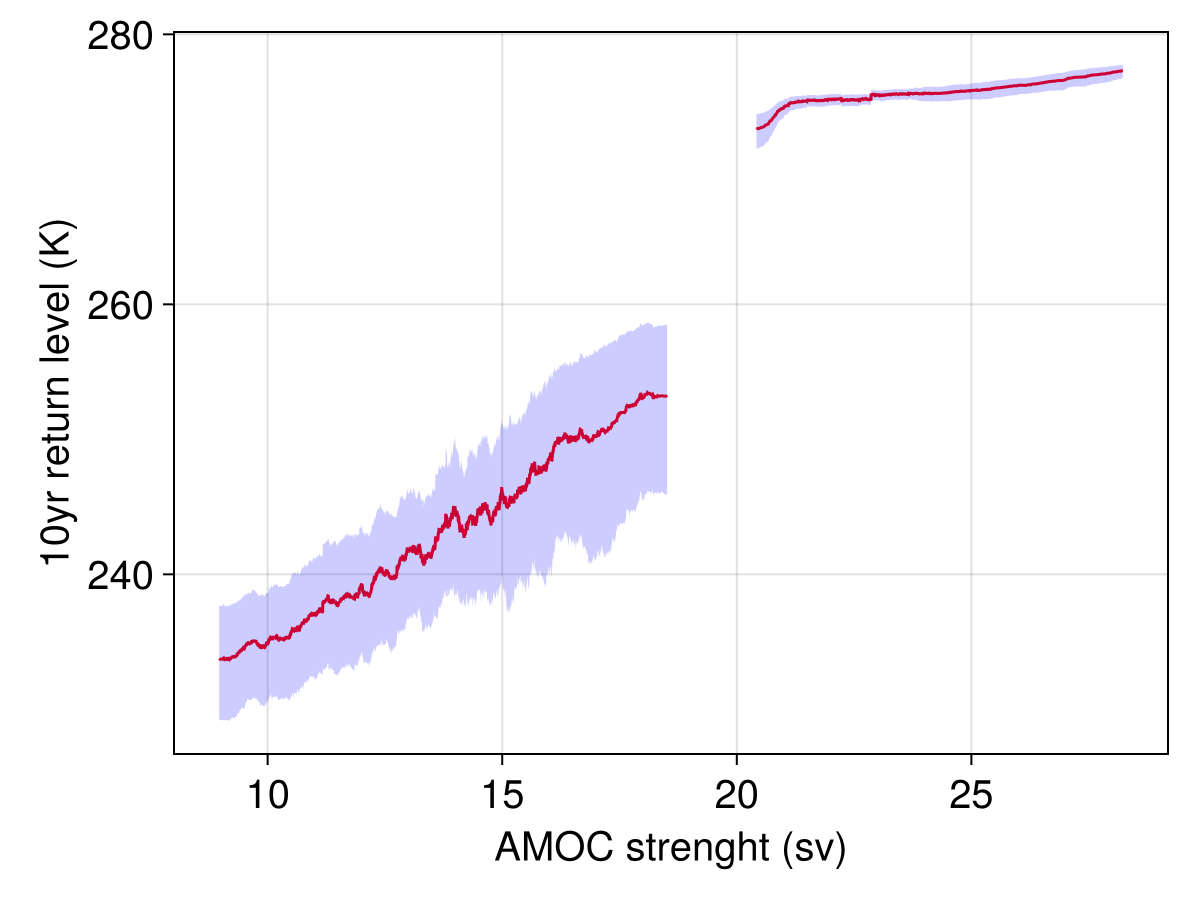}
    \caption{10 year return levels for the analogous climate to the pre-industrial De Bilt chosen in our system.}
    \label{fig:extrapolation_analogous}
\end{figure}

A location that meets this criteria in our simulation is the Atlantic coast of France. In this location we obtain very similar numbers in the return level, if we consider comparing the interstadial state (10 year return levels between $2^\circ$C and 4$^\circ$C depending on AMOC strength) with the stadial state (10 year return levels between $-21^\circ$C and $-40^\circ$C  depending on AMOC strength), see Figure~\ref{fig:extrapolation_analogous}. 

However choosing a similar climate might be subject to cherry-picking, since we are looking for similar results in a vast pool of data. We believe that it is however interesting that we can find a LGM conditions location that shows a similar behaviour in terms of extremes as a PI climate, and carry on that through a range of AMOC strengths in a rapidly changing climate. This is unlikely to be possible to make in general for very different states, or for an arbitrary location.

\section{Discussion and Conclusions}


The linear model for the extremes is useful for pattern detection although it has important biases. Nevertheless paints a spatial landscape of correlations that allow to attribute/associate changes in the parameters to physical processes and thus aid understanding. Its simplicity brings the advantage that it is a model that can be applied to any data series, but comes with a price you pay in accuracy, since most parameters still may have a non-linear dependency on the covariate. This gives rise to errors and bias.  

The non-parametric model has the opposite properties. It accurately shows how the parameters of the GEV distribution change in each location, but potentially it has a different behaviour in each location and shows a variety of them. This method can be used to map extreme value distributions into a covariate that exhibits tipping behaviour, allowing extreme events to be described in a system that transitions between competing states. The results may depend on the suitability of the covariate chosen.

In many locations there is a approximately linear or constant behaviour of the parameters against the AMOC strength within each state, and a jump or non-linear oscillation between the states. This is caused by the transition being very fast, thus the statistics encompass data points coming from the two different states. The estimation of the parameters during the transition is not really reliable for this reason. The hypothesis of approximate stationarity is unlikely to hold.

Studying the extremes through the parameters of the GEV distribution conveys information about the magnitude and frequency of the extremes in several different ways. This is particularly important for a complex system because many different kinds of variability are possible, and in our case many are realised in our simulation. Moreover, some changes to the parameters can be related to the proxy record and to physical processes that may be drivers of said changes. 

Of particular interest are the changes observed in the shape parameter for several reasons. On one hand, because it is traditionally left out of the non-stationary analysis due to its difficulty to estimate, and thus it is hard to obtain sufficient statistical evidence of its changes. 
On the other hand, because it is the dominant parameter that describes the decay of the probability as events get more intense and dictates the shape of the distribution. It is thus an important result that we observe strong and significant changes in the shape parameter in some locations, sometimes linear and sometimes non-linear. At least one other study that used other methodology (run climate to equilibrium and then compare distributions) have found significant but very small changes in the shape parameter of cold extremes for an abrupt CO$_2$ increase scenario \cite{huang2016estimating}. Here we observe strong changes and even changes of sign, which transform the GEV distribution from a finite right-endpoint distribution to a infinite right-endpoint heavy tailed distribution or vice-versa.

The GEV distribution becomes fat-tailed when the value of the shape parameter value is greater than $1/2$. 
To our knowledge this is first evidence of fat-tailed distributions applying to temperature extremes, which typically follow a GEV distribution with a finite right endpoint ($\xi<0$) \cite{krakauer2024normal,huang2016estimating}. An important message from Figure~\ref{fig:fat_tails} is that it shows how a quasi-Gaussian distribution for the temperature minima changes into a fat-tailed distribution by the mediation of a physical process (the retreat of the sea ice).

There are important teleconnections that show up in the temperature extremes that can be related to both physical processes and the proxy record \cite{pontes2024weakening,anderson2026shifting,ditlevsen1999observation,gersonde2003last,verma2023variation,romero2015high}. The model employed here agrees mostly with the proxy record of the Northern Hemisphere, and shows important parallelisms with some sediment cores of the southern Oceans. Since the data of the sediment cores depends on their exact location and in models the patterns vary from model to model \cite{kageyama2013climatic} it is hard to find a direct comparison between the output of our model and the ocean sediment proxy record. There is however evidence of the underlying mechanism of the current changes inducing temperature oscillations near the areas identified in our model.

Climate states can have very different extremal behaviours. Extrapolating even between similar climate states needs to pay attention to the physical processes that occur on them. While we cannot make meaningful comparisons between locations that have different dynamics in different climate states we can pay attention to how physical processes alter the distributions of extremes to obtain information. It is likely however that this information is heavily dependant on context (climate state and model), since large scale physical processes such as AMOC collapse behave differently in models and states.

\subsubsection*{Acknowledgements}

We want to thank several people that have contributed through data and discussions to the content of this paper. Guido Vettoretti, which run the climate simulations and helped solving some practical questions. Anna V. Cutmore, who compiled papers and details of the marine sediment cores. Also to Gabriel Pontes, René van Westen, Irene Malmierca and Marisa Montoya for interesting discussions that have improved the quality of the paper. 

The Past to Future (P2F) project has received funding from the European Union’s Horizon Europe research and innovation programme under grant agreement No.101184070: Funded by the European Union. Views and opinions expressed are however those of the authors only and do not necessarily reflect those of the European Union or the European Climate, Infrastructure and Environment Executive Agency (CINEA). Neither the European Union nor the granting authority can be held responsible for
them.

\bibliographystyle{unsrt}
\bibliography{bib}
\appendix

\section{Algorithm to divide data series}\label{ap:division_algorithm}
Dividing a noisy data series into different states is a process that can be very subjective. This is specially true when states are very short or they shift, the transitions are slow or the noise is very strong. To introduce a measure of objectivity, we employ the following method.

Say we have a stochastic process $X_i$ which starts in the down state. First we define two threshold values $t_{up}$ and $t_{down}$. A transition up$\rightarrow$down only takes place when the process crosses both thresholds, first $t_{down}$ and then $t_{up}$. The points which are considered as belonging to the transition are the ones that lay between the two thresholds. Note that if the process crosses several times the threshold $t_{down}$ without reaching $t_{up}$, all those data points are considered part of the down state, and only the ones beyond the last crossing of $t_{down}$ before $t_{up}$ are considered as part of the transition. Once $t_{up}$ is reached, all points beyond are considered up state. Then the roles of the thresholds reverse and until $t_{down}$ is reached again, the points will be considered up state and only the points between the last $t_{up}$ crossing and the $t_{down}$ crossing are considered part of the transition up$\rightarrow$down.

In the data series of the AMOC in which we have employed the algorithm the thresholds have been chosen to be $t_{down} = 12$ Sv and $t_{up}= 23.5$ Sv. Dividing the data series of the AMOC is hard for some of the reasons stated before. On one hand the states show some form of relaxation. This means that a high $t_{up}$ or a low $t_{down}$ will cut short some of the states and make very long transitions, specially in the interstadial$\rightarrow$stadial transition. On the other hand the transition stadial$\rightarrow$interstadial shows a high amplitude oscillation that can only be captured as a transition by a high $t_{down}$ and a low $t_{up}$. The choice of the thresholds tries to find a compromise that leaves enough points in both states, at the expense of leaving some points that could be considered as part of the transition in the interstadial state, see Figure~\ref{fig:division_figure}.

\section{Linear model for pattern detection}\label{ap:linear_model}

\subsection{Choosing a model}

For a given data series, 8 different models are possible. The stationary model, $\mathcal{M}_{\emptyset}$, is characterized by $\mu_1,\sigma_1,\xi_1\approx 0$. The models where only one parameter changes significantly, 
$\mathcal{M}_{\{\mu\}}$, $\mathcal{M}_{\{\sigma\}}$, and $\mathcal{M}_{\{\xi\}}$, are characterized by  $\sigma_1,\xi_1\approx 0$, $\mu_1,\xi_1\approx 0$, and $\mu_1,\sigma_1\approx 0$, respectively. Similarly defined, the models where two parameters change are $\mathcal{M}_{\{\mu,\sigma\}}$, $\mathcal{M}_{\{\mu,\xi\}}$, and $\mathcal{M}_{\{\sigma,\xi\}}$; and the full non-stationary model is $\mathcal{M}_{\{\mu,\sigma,\xi\}}$. The test described in Reference \cite{coles2001introduction} allows us to compare models that are nested. We say a model $\mathcal{M}_A$ is nested in a model $\mathcal{M}_B$ if the latter is a generalization of the former. In our notation, it translates into the inclusion relation; thus, $\mathcal{M}_A$ is nested in a model $\mathcal{M}_B$ if $A\subset B$. This induces a partial order in the space of models, since we cannot test every pair of models, e.g. $\mathcal{M}_{\{\mu,\sigma\}}$ and $\mathcal{M}_{\{\mu,\xi\}}$.

Choosing the most adequate model for data series cannot be done in a totally objective way via these tests, since we may need a criterion to decide between two non-nested models.

In the literature \cite{robin2020nonstationary,coles2001introduction} it is usually considered unrealistic to observe changes in the shape parameter. 
We give prominence to the ``simpler'' parameters: Preferring models where $\mu$ changes rather than models where $\sigma$ or $\xi$ change. This is because the shape parameter is notoriously hard to estimate with precision \cite{coles2001introduction}, and thus it is usually left as stationary in other studies of non-stationary extremes \cite{robin2020nonstationary}. Similarly other studies consider that in principle there is no physical mechanism for $\sigma$ and $\xi$ to change under a climate slow drift produced by different CO$_2$ concentrations \cite{huang2016estimating}.

Despite this, here we will still consider the shape parameter changes as a feasible possibility due to the strong changes that the system that we study exhibits. While the systems considered in \cite{coles2001introduction,robin2020nonstationary} show a gradual change of trend, the system at hand here shows switching between two competing alternative states with possibly different dynamics.

The statistical test for the non-stationarity of the extremes takes a cautionary approach. If $A$ and $B$ are such that $A\subset B$, one starts with the simplest model $\mathcal{M}_A$ and test if we have enough evidence to assume that more complex model $\mathcal{M}_B$ \cite{coles2001introduction}. 

In light of this arguments, we follow the cautionary approach and set our order of preference among models in the following way:
\begin{equation}\label{eq:simplicity}
    \mathcal{M}_{\emptyset}<\mathcal{M}_{\{\mu\}}<\mathcal{M}_{\{\sigma\}}<\mathcal{M}_{\{\xi\}}<\mathcal{M}_{\{\mu,\sigma\}}<\mathcal{M}_{\{\mu,\xi\}}<\mathcal{M}_{\{\sigma,\xi\}}<\mathcal{M}_{\{\mu,\sigma,\xi\}}
\end{equation}
Whenever available, a test to decide between the models is performed, and preference between tests is given by Eq.\ref{eq:simplicity}. A whole description of the process is given by following sequence:
\begin{enumerate}
    \item Compute all models that converge by a maximum likelihood estimation method. 
    \item Compute a $99\%$ confidence interval for the parameters. If zero belongs to confidence interval of $\mu_1$, $\sigma_1$ or $\xi_1$, discard the model.
    \item Starting from the most simple remaining model according to Eq.\eqref{eq:simplicity}, perform a non-stationarity test with the next nested most simple model. 
    \item Repeat until there are no next model above or all nested models above are rejected by the test.    
\end{enumerate}

Note that this way of selecting a model does not necessarily test all the possible models, but will always produce a model as long as one model converged. For example if only $\mathcal{M}_{\{\mu\}}$, $\mathcal{M}_{\{\sigma\}}$, $\mathcal{M}_{\{\xi\}}$ and $\mathcal{M}_{\{\mu,\sigma\}}$ are available after steps 1 and 2, this procedure will only test $\mathcal{M}_{\{\mu\}}$ against $\mathcal{M}_{\{\mu,\sigma\}}$, and the result of this test will be the output, without considering the also possible test $\mathcal{M}_{\{\sigma\}}$ against $\mathcal{M}_{\{\mu,\sigma\}}$ and ignoring the non-nested model $\mathcal{M}_{\{\xi\}}$ because of our preference of models. This rises the possibility of biases. 

\subsection{Biases}

The statistical method to find trends with a linear model has a series of biases that are of interest for the interpretation of the results of the paper and future research.


There are several sources of bias for this method. One is the convergence and preference order explained in the previous appendix. Also, some forms of non-linearity can mask a trend so that it is not-detected due to opposite and nearly symmetrical cancellations. That can also produce that an opposite trend can be detected sometimes. For example, for some data series in the North Atlantic the scale parameter non-linear model shows that the stadial period has higher variability, but the trend detected through $\sigma_1$ is positive, indicating that variability should be higher during the interstadial. This means that the non-linear increase in variability observed during the transition between states has a higher impact in the linear trend that the lower value of the variability during the interstadial state.

\begin{figure}
    \centering
    \includegraphics[width=\linewidth]{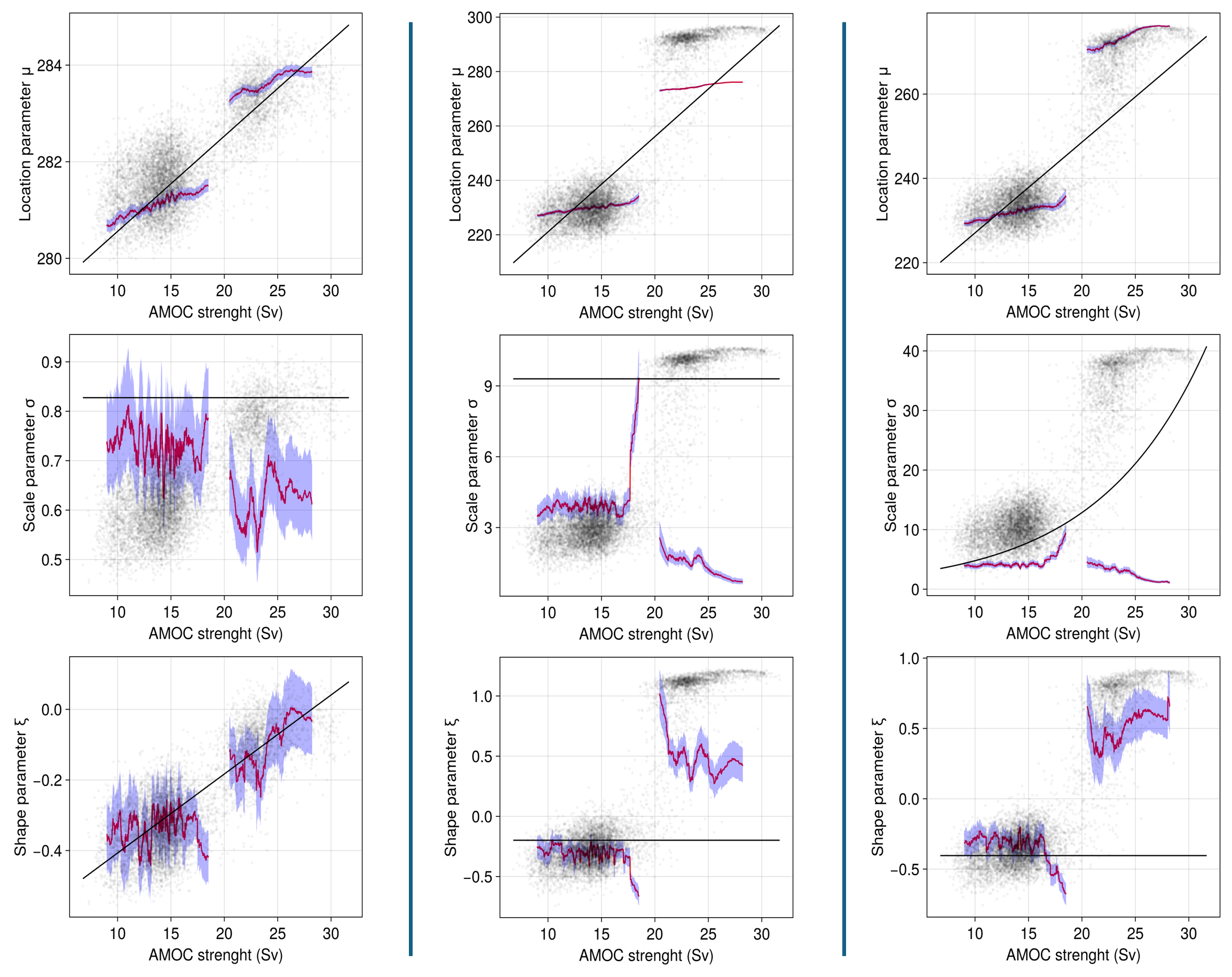}
    \caption{The left panel show a data series corresponding to the Mediterranean region. For this data series no trend in $\sigma$ is detected. The middle and right panel show the data series corresponding two contiguous locations in the North Atlantic area. The tail parameter is not detected to change in these data series, meanwhile the scale parameter is only detected to change in one of them. The change detected in the scale parameter is of the opposite sign, indicating a higher value of $\sigma$ in the interstadial.}
    \label{fig:bias_plots}
\end{figure}

Figure \ref{fig:bias_plots} shows three locations for which the outcome of the linear model does not match the expectations. We overlay the linear model estimate of the dependency on top of the non-parametric fit to facilitate comparison. For the data series on the left, no trend in $\sigma$ is detected although it is easy to see that it is lower during the interstadial. The two other data series are very similar because they correspond to neighbouring locations in the North Atlantic. For these series the tail parameter is not detected to change, even though is in one of the regions were the strongest changes take place. The $\sigma$ parameter is only detected to change in one of them, and it is detected to change with an upward trend, which is the opposite of what we see, since $\sigma$ is lower during the interstadial.

We believe that this biases are responsible for the lack of signal of the parameter $\xi$ in the North Atlantic, which is widespread in the region (see Figure~\ref{fig:planetplot}). Another important bias takes place in the same region but for the parameter $\sigma$, since this parameter is sometimes fitted with a curve that shows the opposite trend as the data, possibly because of the data corresponding to the transition between the states. 

\subsection{Validating}

\begin{figure}
    \centering
    \includegraphics[width=\linewidth]{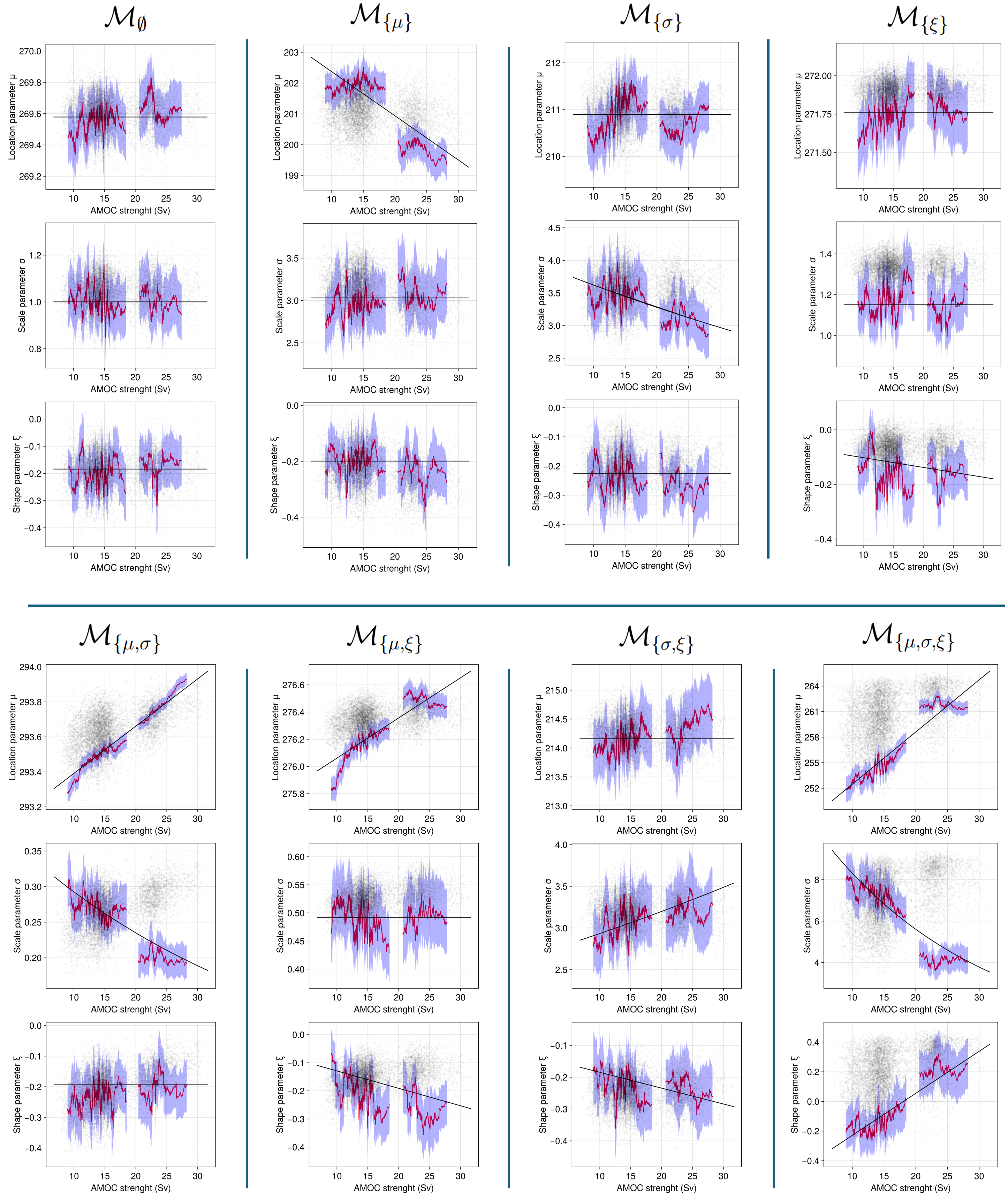}
    \caption{The eight possible models are realised for the data series corresponding to some coordinates in the data set. }
    \label{fig:validation_plots}
\end{figure}

Although the linear model is hard to validate for a large set of data because of the biases explained above, the data is rich enough to show plausible fits to all the possible 8 models. Above we have seen examples of data series that fit best to a stationary model $\mathcal{M}_{\emptyset}$ (inland Africa), the non-stationary only in $\mu$ model $\mathcal{M}_{\{\mu\}}$ (Siberia), non-stationary in $\mu$ and $\sigma$ model $\mathcal{M}_{\{\mu,\sigma\}}$ (Alaska) and the fully non-stationary model $\mathcal{M}_{\{\mu,\sigma,\xi\}}$ (Middle East). Examples of the remaining models can be found in some other locations, see Figure~\ref{fig:validation_plots}. 

Some of the models in the figure show little variation in some parameters, specially when $\mu$ is stationary. The areas where $\mu$ is not detected to be stationary have very little change in general. Changes in the distribution of temperatures are likely to bring a change in the location parameter first and foremost, thus areas without significant trends on $\mu$ are likely to have at most subtle trends on $\sigma$ and/or $\xi$. An example of this is model $\mathcal{M}_{\{\sigma,\xi\}}$ in the figure. Significant trends in $\sigma$ and $\xi$ are detected and visually a trend in $\mu$ is apparent, but looking at the $y$ axis of the plot we see that the shift of the $\mu$ parameter is about $0.5$ degrees K, thus very likely 0 is within the confidence interval of $\mu_1$ and thus the non-stationarity of $\mu$ is discarded.

\end{document}